\documentclass[twocolumn,showpacs,preprintnumbers,amsmath,amssymb,nofootinbib]{revtex4-2}
\usepackage{comment} 
\usepackage{cancel}
\usepackage{graphicx}% Include figure files
\usepackage{dcolumn}% Align table columns on decimal point
\usepackage{bm}% bold math
\usepackage[compact]{titlesec}
\usepackage{float}
\makeatletter
\DeclareRobustCommand{\cev}[1]{%
  {\mathpalette\do@cev{#1}}%
}
\newcommand{\do@cev}[2]{%
  \vbox{\offinterlineskip
    \sbox\z@{$\m@th#1 x$}%
    \ialign{##\cr
      \hidewidth\reflectbox{$\m@th#1\vec{}\mkern4mu$}\hidewidth\cr
      \noalign{\kern-\ht\z@}
      $\m@th#1#2$\cr
    }%
  }%
}
\makeatother
\usepackage{xcolor}

\usepackage{hyperref}
\renewcommand{\arraystretch}{2.6}       % spacing btwn the rows of a non-eqn array
\newcommand{\half}{{{\textstyle\frac{1}{2}}}}
\newcommand{\quarter}{{{\textstyle\frac{1}{4}}}}
\newcommand{\be}{\begin{equation}}
\newcommand{\ee}{\end{equation} }
\newcommand{\beqa}{\begin{eqnarray} }
\newcommand{\eeqa}{\end{eqnarray} }
\newcommand{\ba}{\begin{array}}
\newcommand{\ea}{\end{array}}
\newcommand{\bpm}{\begin{pmatrix}}
\newcommand{\epm}{\end{pmatrix}}

\newcommand{\dis}{\displaystyle}

\newcommand{\rmd}{{\rm d}}
\newcommand{\rd}{{\rmd}}

\newcommand{\ODD}{\mathbf{O}(D,D)}

\newcommand\To{T_{\scriptscriptstyle{{(0)}}}}

\newcommand\Tr{{\scalebox{0.9}{${\mathrm{Tr}}$}}}

\newcommand\cF{{\cal F}}

\newcommand\cH{{\cal H}}

\newcommand\cL{{\cal L}}

\def\na{\nabla}

\newcommand{\rhonew}{\varepsilon}
\newcommand{\rhonewo}{\varepsilon_{0}}
\newcommand{\rhonewzero}{\varepsilon_{0}^{w=0}}
\newcommand{\rhonewthird}{\varepsilon_{0}^{{w=1/3}}}
\newcommand{\pnew}{\mathfrak{p}}
\newcommand{\hh}{\mathfrak{h}}

\newcommand\rr{{\tilde{r}}}

\newcommand{\trd}{{\bigtriangledown}}

\newcommand{\betappn}{\beta_{{{{\rm{PPN}}}}}}
\newcommand{\gammappn}{\gamma_{{{\rm{PPN}}}}}
\newcommand{\GN}{G_{{{{{\rm{N}}}}}}}

\begin{document}

%\preprint{APS/123-QED}
\title{Late-time Cosmology  without Dark Sector but with Closed String Massless Sector}
%%Late-time Open Universe without Dark Sector but with Closed String Massless Sector}
%%Cosmological Feasibility  of  the Closed String Massless Sector up to the Redshift $z=9$\\
%{Supernova, Quasar and String Theory:  %\\
%%  Closed String Massless Sector:\\
%Accelerating  Open Universe   in String Frame}}
%%\title{%Positive pressure accelerates the expansion of the Universe in string frame: \\
%%Type Ia Supernovae, Open Universe, and Closed String Massless Sector}
%%: Accelerating Open Universe  naturally in String Frame}

\author{Hocheol Lee}
\author{Jeong-Hyuck Park}%\email{park@sogang.ac.kr}
\author{Liliana Velasco-Sevilla}
\affiliation{Department of Physics \& Center for Quantum Spacetime, Sogang University, 35 Baekbeom-ro, Mapo-gu, Seoul 04107,  Korea}

\author{Lu Yin}
\affiliation{%Asia  Pacific Center for Theoretical Physics,  Postech, Pohang 37673,  Korea\\
 Department of Physics, Shanghai University, Shanghai, 200444, China} 
% yinlu@shu.edu.cn

\begin{abstract}
\centering\begin{minipage}{\dimexpr\paperwidth-6.3cm}
\noindent  We explore the possibility of  solving   the dark energy and the coincidence problems by postulating  the massless sector of closed strings. This sector constitutes the gravitational multiplet of string theory and,  when applied to  four-dimensional cosmology,  predicts  that \textit{the expansion of an open  Universe defined in  string frame is  readily  accelerating}.  We confront the prediction with the late-time  cosmological  data of  Type Ia supernovae and  quasar absorption spectrum,  which probe the   evolutions  of the Hubble parameter and possibly the fine-structure constant. We report that these   observations are in admirable  agreement with the prediction  without any dark sector or  coincidence problem. 
We estimate   the Hubble constant, $H_{0}\simeq 71.2\pm 0.2\,\mathrm{km/s/Mpc}$. 
\end{minipage}  
%PRL abstract of no more than 600 characters
\end{abstract}

%\pacs{11.25.-w}% PACS, the Physics and Astronomy
                             
%%%
%%11.25.-w 	Strings and branes                             
\maketitle

\section{Introduction} 
The  dark energy  and the  coincidence problems  are  notable challenges that pose a fundamental question to our comprehension of general relativity (GR) at the cosmological level~\cite{Weinberg:1988cp,
Sahni:2004ai,Polchinski:2006gy,Bousso:2007gp,Martin:2012bt,Burgess:2013ara,Padilla:2015aaa,Velten:2014nra,Wang:2016lxa,Bernardo:2022cck}.  While in GR the only gravitational  field is the metric $g_{\mu\nu}$,  in string theory the metric accompanies two additional fields: a two-form potential $B_{\mu\nu}$ and a scalar dilaton $\phi$. The trio  $\left\{g_{\mu\nu},B_{\mu\nu},\phi\right\}$ forms   a massless sector, \textit{i.e.~}the vibrations of closed strings with zero mass.  Their   low-energy effective  action is given by
\be
\!\displaystyle{\int}\rmd^{D}x\,\sqrt{-g}e^{-2\phi}\!
\left[R+4\partial_{\mu}\phi\partial^{\mu}\phi-\textstyle{\frac{1}{12}}H_{\lambda\mu\nu}H^{\lambda\mu\nu}-2\Lambda\right]\,.
\label{Reff}
\ee
Here $H_{\lambda\mu\nu}$ is the three-form field strength of the $B$-field, dubbed $H$-flux, and $\Lambda$ is a possible cosmological constant which may  arise in non-critical string theory or through critical string compactification.

Remarkably, this  $D$-dimensional spacetime action exhibits the $\ODD$ symmetry of T-duality  which transforms the trio $\{g_{\mu\nu},B_{\mu\nu},\phi\}$ into one another~\cite{Buscher:1987sk,Buscher:1987qj,Giveon:1988tt,Duff:1989tf,Tseytlin:1990nb,
Tseytlin:1990va,Gasperini:1991ak}.    More precisely, the metric and the $B$-field form an $\ODD$ covariant multiplet, often called \textit{generalised metric} $\cH_{MN}$,  while the dilaton $\phi$  becomes  an $\ODD$ singlet $d$ in combination with  the determinant of the metric. Through exponentiation it becomes  a scalar density with unit weight:
\be
\ba{ll}
e^{-2d}=\sqrt{-g}e^{-2\phi}\,,~~&~
\cH_{AB}= \left(\ba{cc}
~~g^{-1} & - g^{-1}B \\
Bg^{-1} &~~ g - Bg^{-1}B
\ea\right).
\ea
\label{cHd}
\ee
The generalised metric carries a pair of $\ODD$ vector indices (capital letters, $A,B$), such that under $\ODD$ rotations, the metric and the $B$-field  transform to each other. Further, since $d$ is an $\ODD$ singlet, the dilaton $\phi$ transforms to  the others too.   In particular,  $e^{-2d}$ is the  $\ODD$-symmetric   integral measure (\textit{i.e.~}scalar density with unit weight) of DFT   and  the $\Lambda$-term in (\ref{Reff}) is the unique $\ODD$-symmetric cosmological constant term.\\

\subsection{Prediction from Double Field Theory}
 The $\ODD$ symmetry is surely a nontrivial property of the action~(\ref{Reff}) that assumes  the mathematical  framework of    `Riemannian geometry'.  However, the symmetry becomes manifest in the alternative framework called  double field theory (DFT) which reformulates the action (\ref{Reff}) in terms of the generalised metric and the $\ODD$ singlet dilaton $d$~\cite{Siegel:1993xq,Siegel:1993th,Hull:2009mi,Hull:2009zb,Hohm:2010jy,Hohm:2010pp}.   Over the last decade,  DFT has evolved, further  to possess its own autonomous structure. 

\textit{i)}  First of all, the generalised metric $\cH_{AB}$ and the $\ODD$ singlet dilaton $d$ themselves are the fundamental geometric quantities of  DFT  which have their own definitions  more general  than the right hand sides of the equalities in (\ref{cHd}), \textit{c.f.~}\cite{Lee:2013hma,Ko:2015rha,Morand:2017fnv}.  

\textit{ii)}  In an analogous manner to GR, imposing compatibility with the fundamental fields,  
$\na_{C}\cH_{AB}=0=\na_{A}d$, along with some torsionless conditions,  $\ODD$ symmetric,  DFT-versions of the Christoffel symbol,    covariant derivatives, and scalar/Ricci curvatures were all constructed~\cite{Jeon:2010rw,Jeon:2011cn,Jeon:2011vx}.  In particular,       the first  three  terms in (\ref{Reff})    correspond to  fragments  of   the DFT scalar curvature with the parametrisation~(\ref{cHd}). This suggests that     the action~(\ref{Reff})  should  correspond to   stringy  `pure' gravity that  gravitises  the entire closed string massless sector.  

\textit{iii)}  The $\ODD$-symmetric covariant derivatives then realise the notion of minimal coupling of $\{\cH_{AB},d\}$ to other sectors in string theory~\cite{Jeon:2011vx,Jeon:2012kd,Jeon:2011sq,Jeon:2012hp},   or  to   generic   `matter' such as the Standard Model of  particle physics~\cite{Choi:2015bga}. 

\textit{iv)} Most centrally,  DFT has its own   Einstein equation, namely Einstein Double Field Equation~\cite{Angus:2018mep},\footnote{In our convention, the Newton constant $\GN$ does not appear in the action~(\ref{Reff}), as it is absorbed into the energy-momentum tensor.   It can be  restored   for the gravitational mass, see \textit{e.g.~}(\ref{PPNmetric}) and  (\ref{NewtonMass}).}
\be
G_{AB}=T_{AB}\,.
\label{EDFEODD}
\ee 
Again parallel to GR, but  through the aforementioned $\ODD$ symmetric differential geometry,    Eq.(\ref{EDFEODD}) consists of  DFT-versions of Einstein curvature  on its left hand side and energy-momentum tensor on the right, which   are conserved off-shell   and on-shell  respectively, ${\na_{A}G^{A}{}_{B}=0=\na_{A}T^{A}{}_{B}}$. Through this master formula where $G_{AB}$ comprises the stringy  geometric data, $\{\cH_{AB}, d\}$,  matter \textit{i.e.~}$T_{AB}$ tells stringy geometry  ``how to curve''.

While we refer to  \cite{Angus:2018mep} (especially section 2 therein) for detailed  explanation  of the above $\ODD$ symmetric formalism,    we shall not directly utilize it   in the present paper. Hereafter the mathematical language as for covariant derivatives and curvatures is to be (familiar) Riemannian.  To start,   adopting the parametrisation~(\ref{cHd})  for the DFT fundamental fields, the   single master  formula~(\ref{EDFEODD})  reduces to  the three equations of motion of the metric, the $B$-field, and the dilaton~\cite{Angus:2018mep},
\be
\ba{rll}
R_{\mu\nu}+2\trd_{\mu}(\partial_{\nu}\phi)-\quarter H_{\mu\rho\sigma}H_{\nu}{}^{\rho\sigma}
&\!=&\! K_{(\mu\nu)}\,,\\
\half e^{2\phi}\trd^{\rho}\!\left(e^{-2\phi}H_{\rho\mu\nu}\right)&\!=&\! K_{[\mu\nu]}\,,\\
R+4\Box\phi-4\partial_{\mu}\phi\partial^{\mu}\phi-\textstyle{\frac{1}{12}}H_{\lambda\mu\nu}H^{\lambda\mu\nu}&\!=&\!\To\,.
\ea
\label{EDFE}
\ee
While  the quantities appearing on the right hand sides of the equalities are  \textit{a priori}  various components of the $\ODD$ symmetric  energy-momentum tensor $T_{AB}$,  they can be   alternatively obtained  from the variation of the  matter Lagrangian, $\cL_{\rm matter}$ coupled to (\ref{Reff}):
\be
\ba{rrr}
K^{(\mu\nu)}&=&\left.e^{2d}\frac{\delta\cL_{\rm matter}}{\delta g_{\mu\nu}}\right|_{B\,\&\, d~{\rm{fixed}}}\,,\\
K^{[\mu\nu]}&=&-\left.e^{2d}\frac{\delta\cL_{\rm matter}}{\delta B_{\mu\nu}}\right|_{d\,\&\,g~{\rm{fixed}}}\,,\\
T_{(0)}&=&\left.\frac{1}{2}e^{2d\,}\frac{\delta\cL_{\rm matter}}{\delta d}\right|_{g\,\&\,B~{\rm{fixed}}}\,.
\ea
\label{KKT}
\ee
We stress that the independent  fields  in the above functional partial derivatives are  the metric, $B$-field, and the $\ODD$ singlet dilaton,
\be
d=\phi-\frac{1}{2}\ln\sqrt{-g}\,.
\ee 
Instead, if we chose $\{g,B,\phi\}$, rather than  $\{g,B,d\}$, as independent variables, we would have the conventional  energy-momentum tensor comprising $K^{(\mu\nu)}$ and $\To$,
\be
\scalebox{1}{${T}^{\mu\nu}=\left.\frac{1}{\sqrt{-g}}\frac{\delta\cL_{\rm matter}}{\delta g_{\mu\nu}}\right|_{B\,\&\, \phi~{\rm{fixed}}}=e^{-2\phi}\Big[K^{(\mu\nu)}-\frac{1}{2}g^{\mu\nu}\To\Big]$}\,,
\label{conT}
\ee
such that
\be
K_{(\mu\nu)}=e^{2\phi}T_{\mu\nu}+\frac{1}{2}g_{\mu\nu}\To\,,
\label{KTT}
\ee
and, from (\ref{EDFE}) on-shell, $T_{\mu\nu}$ would match  the ordinary   Einstein curvature up to the dilaton and $H$-flux.  \\

The effective action~(\ref{Reff}) is written in string frame and if we perform a field redefinition of the metric,
\be
g_{\mu\nu}=e^{\frac{4\phi}{D-2}}g^{\rm{E}}_{\mu\nu}\,,
\label{ggE}
\ee
it is possible to rewrite the action in Einstein frame where a standard Einstein--Hilbert term appears without involving the dilaton. For the given action~(\ref{Reff}) alone, this change of a frame should not change any physics.  We proceed to consider  coupling the action to a point particle.  The experimentally well-tested,  Equivalence Principle or the principle of  free-fall motion,    insists on  coupling  the test particle  minimally to a metric.   But, this minimal coupling is not invariant under the change of the frame, as the following two  actions of a  particle are inequivalent,
\be
\dis{\int {\rm{d}}\tau~\sqrt{-g_{\mu\nu}\frac{\rd x^{\mu}}{\rd\tau}\frac{\rd x^{\nu}}{\rd\tau}}\quad\mathit{~vs.~}\quad
\int {\rm{d}}\tau~\sqrt{-g^{\rm{E}}_{\mu\nu}\frac{\rd x^{\mu}}{\rd\tau}\frac{\rd x^{\nu}}{\rd\tau}}\,.}
\label{ptc2}
\ee
In each case, one may  change the frame,  \textit{e.g.~}
\be
\dis{\int {\rm{d}}\tau~\sqrt{-g_{\mu\nu}\frac{\rd x^{\mu}}{\rd\tau}\frac{\rd x^{\nu}}{\rd\tau}}}=
\dis{\int {\rm{d}}\tau~e^{\frac{2\phi}{D-2}}\sqrt{-g^{\rm{E}}_{\mu\nu}\frac{\rd x^{\mu}}{\rd\tau}\frac{\rd x^{\nu}}{\rd\tau}}}\,,
\ee
but the  non-equivalence of the two particle actions in (\ref{ptc2})  persists.  The truth is that, the proper time, the geodesic motion of free-falling, and---as we shall see through (\ref{HEH})---the Hubble parameter are all frame-dependent notions, while they are  physical observables.  In the conventional framework that is based on Riemannian geometry, there seems  no compelling argument how to  fix the dilaton  coupling in order to choose the correct action in  (\ref{ptc2}). 

As stated above,  the  $\ODD$ symmetry principle leads to the $\ODD$ symmetric minimal coupling of $\{\cH_{AB},d\}$  to various types of matter.  Completely through  the $\ODD$-symmetric tensor~$\cH_{AB}$ and scalar density $e^{-2d}$~(\ref{cHd}),  the symmetry   dictates unequivocally how  $\{g_{\mu\nu},B_{\mu\nu},\phi\}$, hence the action~(\ref{Reff}),    should couple to  matter.  In  particular in \cite{Ko:2016dxa},  the   $\ODD$-symmetric   worldline action for a point particle was  constructed   in terms of the generalised metric $\cH_{AB}$. Since the $\ODD$ singlet dilaton, or $e^{-2d}$, carries a nontrivial   diffeomorphism weight, it cannot couple to the  diffeomorphism invariant worldline action.\footnote{As an analogy, multiplying \textit{e.g.~}$\sqrt{-g}$ or $\sqrt{-g^{\rm{E}}}$  to the particle actions in (\ref{ptc2}) would break their target spacetime diffeomorphism invariance and hence is forbidden.}  The  $\ODD$-symmetric particle  action was shown to  reduce---after gauging away half of doubled coordinates~\cite{Park:2013mpa}---to the particle action on the left in (\ref{ptc2}). That is to say, the $\ODD$ symmetry  predicts that the Equivalence Principle of free-falling, free of a fifth force, holds in the string frame rather than in Einstein's.  This result is also consistent with the idea of string theory that matter is made of vibrating tiny strings that couple minimally to the string frame metric  for sure.

On the other hand, when coupled to the Standard Model~\cite{Choi:2015bga}---which is not a particle action but a quantum field theory action---the $\ODD$ symmetry  dictates that,  spinorial fermions (carrying a half-unit weight of  diffeomorphism) interact with the $H$-flux~\cite{Jeon:2011vx}, while gauge bosons do so with the dilaton~\cite{Jeon:2011kp}:
\be
\int\rmd^{4}x~\bar{\psi}\big[i
\gamma^{\mu}{(D_{\mu}}{+\textstyle{\frac{1}{24}}}H_{\mu\nu\rho}\gamma^{\nu\rho})-m\big]\psi-\quarter e^{-2d}\Tr(F_{\mu\nu}F^{\mu\nu})\,,
\label{QED}
\ee
where $D_{\mu}$ is a usual covariant derivative involving  spin/gauge connections.    The quantities~$K_{\mu\nu}, \To$ of (\ref{KKT}) are then   unambiguously determined for each matter content. In particular, from (\ref{KKT}), and (\ref{QED}), in addition to the symmetric $K_{(\mu \nu)}$, spinorial fermions produce  nontrivial $K_{[\mu\nu]}$, while   the cosmological constant  and the gauge bosons (or gluon condensate) contribute to the scalar component of the energy-momentum tensor, %$\To$,
\be
\To=2\Lambda+\frac{1}{4} \Tr(F_{\mu\nu}F^{\mu\nu})\,.
\label{ToLF2}
\ee

In this way, string theory predicts its own gravity that  comprises     the closed string massless sector $\{g_{\mu\nu},B_{\mu\nu},\phi\}$,  modifies GR in a very specific way (the effective action~(\ref{Reff}) and its $\ODD$-symmetric coupling to matter), and thus   calls  for verification. \\

\subsection{Solar System Test}
To test the gravity of string theory,  specifically as described by (\ref{EDFEODD}) or (\ref{EDFE}), on an astrophysical scale, a parametrized post-Newtonian (PPN) analysis was conducted in \cite{Choi:2022srv}, following the methodology outlined in \cite{Will:2014kxa}.   For an isotropic  metric around a spherical object, or ``star'', set in the string frame,
\be
\ba{rll}
\rd s^{2}&=&-\left(1-\frac{2M\GN}{r}+\frac{2\betappn(M\GN)^2}{r^2}+\cdots\right)\rd t^2\\
{}&{}&\!\!\!\!\!\!\!\!\!\!\!\!\!\!\!\!\!\!\!+\left(1+\frac{2\gammappn M\GN}{r}+\cdots\right)\left(\rd r^2+r^2\rd\vartheta^2+r^{2}\sin^{2\!}\vartheta\,\rd\varphi^2\right)\,,%\rd \vec{x}\cdot\rd\vec{x}\,,
\ea
\label{PPNmetric}
\ee
the PPN analysis revealed the following results. Firstly, the Newtonian mass  is given   by the integral of the temporal component of $K_{(\mu\nu)}$~(\ref{KKT}) rather than  $T_{\mu\nu}$~(\ref{conT}) over the star,
\be
M\GN=-\frac{1}{4\pi}\int\rd^{3}x~e^{-2d}K_{t}{}^{t}\,.
\label{NewtonMass}
\ee
Secondly, the  PPN parameter $\betappn$ should be unity,  
\be
{\betappn=1}\,, 
\label{beta}
\ee
as implied by a weak energy condition. This is the same value as in GR, \textit{i.e.~}Schwarzschild geometry, and is consistent with observations. Thirdly,  the other PPN parameter $\gammappn$ is  essentially  a (generalised) equation-of-state parameter, as determined through
\be
\gammappn-1= \frac{\dis{\int\rd^{3}x~e^{-2d}\big(K_{\mu}{}^{\mu}-\To+\frac{1}{6}H_{\lambda\mu\nu}H^{\lambda\mu\nu}\big)}}{\dis{-\int\rd^{3}x~e^{-2d}K_{t}{}^{t}}}\,.
\label{gamma}
\ee
While the denominator is the Newtonian mass of the spherical object times  $4\pi\GN$~(\ref{NewtonMass}),  the numerator corresponds to the volume integral of  the ``Chameleon mass''~\cite{Khoury:2003aq,Brax:2004qh} of the  dilaton, as  we can acquire a Klein--Gordon type equation after subtracting the third  from the trace of the first in (\ref{EDFE})~\cite{Choi:2022srv},
\be
\Box\Big(e^{-2\phi}\Big)=\Big(K_{\mu}{}^{\mu}-\To+\frac{1}{6}H_{\lambda\mu\nu}H^{\lambda\mu\nu}\Big)e^{-2\phi}\,,
\ee
where the quantity inside the parentheses on the right hand side of the equality can be viewed as the  effective or Chameleon mass of the scalar field $e^{-2\phi}$.

The electric $H$-flux is trivial everywhere for the spherical configuration, due to the energy condition underlying (\ref{beta})~\cite{Choi:2022srv}, while the magnetic $H$-flux can be nontrivial only inside the star:
\be
H^{r\vartheta\varphi}=-2e^{2d}\dis{\int_{r}^{r_{\star}}\rd r^{\prime}~e^{-2d}K^{[\vartheta\varphi]}}\,,
\label{mH}
\ee
where $r_{\star}$ is the radius of the star. This integral certainly vanishes outside  the star (${r>r_{\star}}$) where  $K^{[\vartheta\varphi]}$ is trivial. In our normalisation of the energy-momentum tensor~(\ref{EDFE}),    (\ref{NewtonMass}), $K_{\mu\nu}$ and $\To$ are linear order in $\GN$, and thus the $H$-flux squared term in (\ref{gamma}) is  subleading and hence less important than others.  The results of (\ref{NewtonMass}), (\ref{beta}), (\ref{gamma}), and (\ref{mH}) demonstrate that, in contrast to GR where Birkhoff's theorem  holds,   more than one component of the energy-momentum tensor   source the stringy gravity  through (\ref{EDFE}) and determine  the geometry (\ref{PPNmetric}). Therefore, probing the geometry outside the star, it is  possible to examine the equation-of-state of the inner compact object, star.

The current  stringent observational bounds for the solar gravity  are  $\gammappn=1+(2.1\pm 2.3)\times 10^{-5}$ (Shapiro time-delay by Cassini spacecraft)~\cite{Will:2014kxa,Bertotti:2003rm}. Consequently,  the theory   can pass the solar system test, provided  Sun meets
\be
 \left|\,\frac{\dis{\int_{\scriptscriptstyle{\rm{SUN}}}}\rd^{3}x~~e^{-2d}\big(K_{\mu}{}^{\mu}-\To+\frac{1}{6}H_{\lambda\mu\nu}H^{\lambda\mu\nu}\big)}{\dis{\int_{\scriptscriptstyle{\rm{SUN}}}}\rd^{3}x~~e^{-2d}K_{t}{}^{t}}\,\right|~\lesssim~10^{-5}\,.
\label{SolarTest}
\ee
That is to say,  the dilatonic Chameleon mass  should be  $10^{-5}$ times smaller than the Sun's Newtonian mass.  
The Einstein Double Field Equation~(\ref{EDFEODD})  alone, or its expanded version~(\ref{EDFE}),  cannot determine whether this inequality holds or not. The solar system test of DFT remains inconclusive. 

On the one hand, non-relativistic (featureless) point particles fail to meet the  inequality~(\ref{SolarTest}). They set $\To$ and $K_{[\mu\nu]}$, hence $H$-flux~(\ref{mH}), all to trivial values. Equation~(\ref{KTT}) then implies that $K_{(\mu\nu)}$ is proportional to the ordinary energy-momentum tensor, $K_{(\mu\nu)} = e^{2\phi}T_{\mu\nu}$, and the PPN parameter $\gammappn$ amounts to three times the ordinary equation-of-state parameter, \textit{i.e.~}$\gammappn = 3w$. Non-relativistic cold matter particles would have negligible $w$, hence $\gammappn \approx 0$ in conflict with the observation.

On the other hand,  the inequality can be satisfied at a subhadronic  level  by pseudo-conformal matter such as massless gluons or massless chiral models (\textit{cf.}~\cite{Lee:2021hrw, Fujimoto:2022ohj, Rho:2022wco}). This may hint at the gravitational form factor inside hadrons \cite{Polyakov:2018zvc, Shanahan:2018nnv, DelDebbio:2021xwu}. A ``pressureless dust" has negligible thermal pressure outside the dust. However, its energy is confined inside, and so should be any intrinsic pressure which differs from the thermal pressure (see \cite{Burkert:2018bqq} for experimental evidence of immense pressure inside protons). \\

\subsection{Objective: Cosmological Test}
This work aims to test the stringy gravity on a cosmological scale,  potentially offering    a   solution to the dark energy and  the coincidence  problems.   We test  the theory, or the cosmological implication of (\ref{EDFE}),    against  two sets of  late-time  cosmological data:  the Pantheon$\mathbf{+}$  sample  of spectroscopically confirmed Type Ia supernovae (SNe Ia)~\cite{Scolnic:2021amr,Riess:2021jrx} and  absorption  spectra of  quasars~\cite{King:2012id,Wilczynska:2015una,Martins:2017qxd,Wilczynska:2020rxx}.  
The former~\cite{Scolnic:2021amr,Riess:2021jrx} consists of  $1701$ data points,  out of  which  $1583$ have the redshift range,  $0.01\leq z \leq 2.26$,  and  provide us a reliable  way to measure the Hubble parameter at different $z\,$ (the  points of $z\leq 0.01$ are  unreliable for the  cosmological fits of the expansion history of the Universe). The latter~\cite{King:2012id,Wilczynska:2015una,Martins:2017qxd,Wilczynska:2020rxx}  consisting    of $199$ points over $0.22\leq z\leq 7.06$ imposes strict limits on the temporal variation of the fine-structure constant $\alpha$ and hence the dilaton $\phi$, since    (\ref{QED}) suggests that $\alpha$ should be  proportional to $e^{2\phi}$~\cite{Murphy:2000pz,Webb:2000mn,Uzan:2002vq,Murphy:2003hw,Damour:1994zq,Damour:2002mi,Damour:2002nv,Chiba:2006xx},
\be
\frac{\alpha(z)}{\alpha(0)}\,=\,\frac{e^{2\phi(z)}}{e^{2\phi(0)}}~\,\stackrel{?}{\approx}~1\quad\mbox{~for~}~~z\,\lesssim\,7\,.
\label{alphaphi}
\ee  
Besides, we shall also estimate      the   Lorentz symmetry breaking  effect of  the $H$-flux  in    (\ref{QED}) and  examine  the inequality of (\ref{SolarTest}).  

The string cosmology of the closed string massless sector has a  longer history than DFT, notably \cite{Gasperini:1991ak,Lidsey:1999mc,Gasperini:2001pc,Gasperini:2007zz}.  Compared to them,    the novelty of the present work lies in the following three  aspects: \textit{i)}  contrast of the string frame against  the Einstein frame including the  natural realization of the  accelerating  expansion of Universe   in the  string frame,  \textit{ii)}  the dominance  of  negative spatial curvature (${k<0}$, \textit{i.e.~}open Universe),  and  \textit{iii)} the test of the model against the real data including a two-parameter fitting of an analytic solution. \\

\section{$\ODD$ Complete  Friedmann Equations\label{SECODD}}
%\section{Accelerating Open Universe    in String Frame}
In supersymmetric or bosonic string theory, the critical  spacetime   dimension is  ${D=10}$ or ${D=26}$ respectively. The process of reducing  from these higher dimensions to the observed four dimensions, \textit{i.e.~}compactification  is  an actively researched topic, see \textit{e.g.~}\cite{Grana:2005jc,Marconnet:2022fmx,Cicoli:2023opf,VanRiet:2023} for recent discussions.  In this work, we simply postulate  a four-dimensional spacetime which exhibits unbroken $\mathbf{O}(4,4)$ symmetry. This assumption ensures that the equations in (\ref{EDFE}) hold, and  their right-hand sides, \textit{i.e.~}$\left\{K_{(\mu\nu)}, K_{[\mu\nu]}, \To\right\}$,  represent generic matter contents which can  accommodate      Kaluza--Klein modes below a  compactification scale, or/and effectively   the physical degrees in the Standard Model~\cite{Choi:2015bga} (\textit{cf.~}non-perturbative  approach~\cite{Brandenberger:1988aj,Brandenberger:2006vv,Brandenberger:2008nx,Brandenberger:2023ver} or non-Riemannian scenario for the internal space~\cite{Morand:2017fnv,Cho:2018alk,Berman:2019izh,Cho:2019ofr,Park:2020ixf,Gallegos:2020egk}).

The  cosmological ansatz of the closed string massless sector  consists of a  Friedmann--Lema\^{i}tre--Robertson--Walker    metric, 
 \be
\!\rd s^2=-N(t)^2\rd t^2+a(t)^2\!\left[\frac{\rd \rr^2}{1-k\rr^2}+\rr^2\!\left(\rd\vartheta^2+\sin^2\!\vartheta\rd\varphi^2\right)\right]\,,
\label{FLRW}
\ee
a time-varying dilaton $\phi(t)$, and a  two-form $B$-field producing a homogeneous and isotropic magnetic $H$-flux,
\be
\!\!{B_{(2)}=\frac{\hh \rr^2\cos\vartheta}{\sqrt{1-k \rr^2}}\rd \rr\wedge\rd\varphi\,,}\quad
{H_{(3)}=\frac{\hh \rr^2\sin\vartheta}{\sqrt{1-k \rr^2}}\rd \rr\wedge\rd\vartheta\wedge\rd\varphi\,,}
 \label{FLRWBH}
 \ee
where $\hh$ should be constant and $\rr$ denotes an areal radius up to the scale factor~$a(t)$, as seen in the metric~(\ref{FLRW}). It can be converted to an isotropic radius $r$ such as in (\ref{PPNmetric}) through the following radial coordinate transformation,
\be
\ba{lll}
\dis{\rr=\frac{r}{1+\frac{1}{4}k{r^2}}}\quad&\qquad\Longleftrightarrow\qquad&\quad \dis{r=\frac{2\rr}{1+\sqrt{1-k\rr^2}}}\,,
\ea
\label{Arrr}
\ee
which gives  as relevant to the $B$-field~(\ref{FLRWBH}), 
\be
\frac{\rr^2 \rd\rr}{\sqrt{1-k\rr^2}}=\frac{r^2 \rd r}{\left(1+\frac{1}{4}k{r^2}\right)^3}\,,
\ee
and to the metric~(\ref{FLRW}),
\be
\frac{\rd\rr^2}{1-k\rr^{2\,}} +\rr^2\big(\rd\vartheta^2+{\sin^2\!\vartheta}\rd\varphi^2\big)=\frac{\rd r^2{+r^2}\big(\rd\vartheta^2+{\sin^2\!\vartheta}\rd\varphi^2\big)}{\left(1+\frac{1}{4}k{r^2}\right)^2}\,.
\label{AIconversion}
\ee

With the cosmological ansatz (\ref{FLRW}) and (\ref{FLRWBH}) plugged in,---regardless of the choice of the radial coordinate---the trio of the Einstein double field equation~(\ref{EDFE})     modify   the Friedmann equations~\cite{Angus:2019bqs},
\be
\ba{l}
\!\!{\frac{2}{N}\frac{\rd \phi}{\rd t}}={3H\pm\sqrt{3H^{2}+2 \rhonew e^{2\phi} +  \To -\frac{6k}{a^{2}}+\frac{\hh^{2}}{2a^{6}}}\,,}\\
\!\!{\frac{1}{N}\frac{\rd H}{\rd t}}=\pnew e^{2\phi}-\frac{2k}{a^{2}}+\frac{\hh^{2}}{2a^{6}}
\pm H\sqrt{\scalebox{0.87}{$3H^{2}+ 2\rhonew e^{2\phi} + \To -\frac{6k}{a^{2}}+\frac{\hh^{2}}{2a^{6}}$}}\,.
\ea
\label{OFE}
\ee
Here  the Hubble parameter is defined, with the lapse function~$N(t)$ and the scale factor~$a(t)$ in the string frame~(\ref{FLRW}), by
\be
H(t)=\frac{\,\rd\ln a(t)}{N(t)\rd t}\,.
\ee
Further,  as for the cosmological fluid-like matter content,    we set a mass density~$\rhonew$ that is responsible for the Newton potential~(\ref{NewtonMass})~\cite{Cho:2019npq,Choi:2022srv} along with   its spatial counterpart~$\pnew$  which we simply call   ``pressure''  henceforth,\footnote{Our identifications  of the mass density and the  ``pressure''  in terms of $K_{\mu\nu}$ and $\To$ differ from those in   \cite{Angus:2019bqs} (see also \cite{Lescano:2021nju}),   but are  motivated by the post-Newtonian analyses of \cite{Cho:2019npq,Choi:2022srv}.   }
\be
\ba{ll}
\rhonew(t)\equiv -e^{-2\phi}K_{t}{}^{t}\,,\qquad&\quad\dis{ \pnew(t)\equiv 
\frac{1}{3}\sum_{i=1}^{3}~e^{-2\phi}K_{i}{}^{i}\,,}
\ea
\label{rhonewpnew}
\ee
which satisfy a continuity  equation involving the scalar component of the energy-momentum tensor~$\To(t)$, 
\be
{\frac{1}{N}\frac{\rd \rhonew}{\rd t}+3 (\rhonew + \pnew)H+ \frac{e^{-2\phi}}{2N}\frac{\rd \To}{\rd t} =0 \,.}
\label{conservation}
\ee
There exist  two branches of the equations distinguished by the upper $+$ and lower $-$ signs,  since    $\frac{2}{N}\frac{\rd \phi}{\rd t}$ in (\ref{OFE}) is actually  an algebraic solution to a quadratic equation,
\be
\frac{\rhonew e^{2\phi}}{3H^2}+\frac{\To}{6H^{2}}-\frac{k}{a^{2}H^{2}}+\frac{\hh^{2}}{12a^{6}H^{2}}=1+\frac{2(\dot{\phi}/N)^{2}}{3H^{2}}-\frac{2\dot{\phi}/N}{H}\,.
\label{ForOmega}
\ee
The left hand side of this equality leads to our identification of    `density parameters': 
\be
\ba{rllrll}
\Omega_{\rhonew}&\equiv&\dis{\frac{\rhonew e^{2\phi}}{3H^2}}\,,\qquad&\qquad
\Omega_{\Lambda}&\equiv&\dis{\frac{\Lambda}{3H^2}}\,,\\
\Omega_{k}&\equiv&\dis{-\frac{k}{a^2H^2}}\,,\qquad&\qquad
\Omega_{\hh}&\equiv&\dis{\frac{\hh^2}{12 a^{6}H^2}}\,.
\ea
\label{defO}
\ee
Note that, hereafter  for simplicity,  we set  ${\To\equiv 2\Lambda}$ rather than the full expression in  (\ref{ToLF2}).  This reduces (\ref{conservation}) to an ordinary continuity equation. By assuming this, we   are ignoring    the   contribution from the gluon condensate  to the scalar component of the energy-momentum tensor,  \textit{i.e.~}$\To$ in (\ref{ToLF2}). This may be justified if the equal number of electric and magnetic gluon fluxes  cancel each other~\cite{DelDebbio:2021xwu,Choi:2022srv}.    Apparently for any time-varying nontrivial dilaton,  Eq.(\ref{ForOmega}) implies that the  sum of the above four density parameters is  not equal to unity.     The time evolution of the dilaton in (\ref{OFE}) can be re-expressed in terms of   the density parameters, 
\be
\frac{2\dot{\phi}/N}{H}=3\pm\sqrt{3+6\big(\Omega_{k}{+\Omega_{\rhonew}}{+\Omega_{\Lambda}}{+\Omega_{\hh}}\big)}\,.
\label{dotphiOmega}
\ee
We continue to  set ${a(t_{0})\equiv 1}$ and ${\phi(t_{0})\equiv 0}$ at present time $t_0$.  Then, for  a  perfect fluid having    an  equation-of-state parameter,
\be
w=\frac{\pnew}{\rhonew}=-\sum_{i=1}^{3}~\frac{K_{i}{}^{i}}{3K_{t}{}^{t}}\,,
\ee
the continuity equation~(\ref{conservation}) with $\To=2\Lambda$  is   solved,  as usual, by
\be
\rhonew(t)={\rhonewo}{a(t)^{-3(1+w)}}\,,
\quad\quad \pnew(t)={w\rhonewo}{a(t)^{-3(1+w)}}\,.
\label{rhonewep}
\ee
After all, as for the $\ODD$-complete Friedmann equations~(\ref{OFE}),  we have maximum six free or fitting  parameters, 
\be
\left\{H_{0}\,,\,k\,,\,\hh\,,\,\Lambda\,,\,\rhonewo\,,\,w\right\}\,,
\ee
which we  shall estimate below  in comparison with observational  data.

Finally, it is worth while to note that if the dilaton $\phi$ were strictly constant  ($\dot{\phi}=0$) and the magnetic $H$-flux vanished ($\hh=0$), the $\ODD$-complete Friedmann equations~(\ref{OFE}) would   reduce to the ordinary Friedmann equations~\cite{Angus:2019bqs}. However, for $\phi$ to remain strictly constant, the right hand side of the equality in the upper equation of (\ref{OFE}) must be  canceled out completely, which would be a redundant constraint and hence we believe very unlikely to occur.  The $\ODD$-complete and ordinary Friedmann equations appear fundamentally distinct.  Having said that, the present paper focuses on the late-time cosmology of the redshift $z\lesssim 7$ over which the dilaton  varies  slowly   converging to a constant in  an open Universe, as we shall  see  in FIG.\,\ref{FIGphiz}, (\ref{analyticphi}), and (\ref{limitphi}). 

%~\\ 

\subsection{Acceleration Readily in String Frame}
It is worth while to eliminate the square-root in (\ref{OFE}) and derive   an expression for the cosmic acceleration,
\be
\ba{l}
\frac{1}{a}\left(\frac{\rd~}{N\rd t}\right)^2a=\frac{1}{N}\frac{\rd H}{\rd t}+H^2\\
=\frac{2}{3}(\dot{\phi}/N)^2-\frac{k}{a^{2}}+\frac{5\hh^2}{12a^{6}}-\frac{1}{6}\To+(\pnew-\frac{1}{3}\rhonew)e^{2\phi}-H^2\,.
\ea
\label{acceleration}
\ee
The deceleration parameter,
\be
{q=-\frac{1}{H^2a}}\left(\frac{\rmd~}{N\rmd t}\right)^{2}a\,,
\ee
is then obtained  in terms of the density parameters,  either from (\ref{OFE})
\be
q=-\left[1{+2}\Omega_{k}{+3w}\Omega_{\rhonew}{+6}\Omega_{\hh}\pm\sqrt{3{+6}\big(\Omega_{k}{+\Omega_{\rhonew}}{+\Omega_{\Lambda}}{+\Omega_{\hh}}\big)}\,\right],
\label{decq}
\ee
or equivalently from (\ref{acceleration}),
\be
q=1-\left[\frac{2(\dot{\phi}/N)^2}{3H^2}+\Omega_{k}+5\Omega_{\hh}-\Omega_{\Lambda}+(3w-1)\Omega_{\rhonew\,}\right]\,.
\label{decq2}
\ee
Remarkably, a feature that was not emphasized in the original derivation~\cite{Angus:2019bqs}  but is now clear is that,  the acceleration is readily  achievable. From (\ref{decq}), firstly for the upper plus sign,  $q$ is obviously  negative as long as
\[
1+2\Omega_{k}+3w\Omega_{\rhonew}+6\Omega_{\hh}>-\sqrt{3+6\big(\Omega_{k}+\Omega_{\rhonew}+\Omega_{\Lambda}+\Omega_{\hh}\big)}\,.
\]
In particular,  when    the equation-of-state parameter~$w$ and the density parameters  are all non-negative, we have $q<0$.  However, in order to  suppress the variation of the dilaton,  in view of the  late-time observational constraint~(\ref{alphaphi}),  we have to choose   the lower minus sign in (\ref{OFE}), (\ref{dotphiOmega}), and (\ref{decq}).  Even so,  the acceleration can  easily occur.  For example,   any of the following cases lets ${q<0}$,\vspace{-5pt}
\be
\ba{l}\vspace{-5pt}
\{\Omega_{k}>1~\,\&\,~\Omega_{\varepsilon}{=\Omega_{\hh}}{=\Omega_{\Lambda}}{=0}\}\,,\\\vspace{-5pt}
\{\Omega_{\hh}>\frac{1}{6}~\,\&\,~\Omega_{\varepsilon}{=\Omega_{k}}{=\Omega_{\Lambda}}{=0}\}\,,\\\vspace{-5pt}
\{-\frac{1}{3}>\Omega_{\Lambda}>-\frac{1}{2}~\,\&\,~\Omega_{\varepsilon}{=\Omega_{k}}{=\Omega_{\hh}}{=0}\}\,,\\\vspace{-5pt}
\{\Omega_{\varepsilon}>0,~~w>\frac{\sqrt{3+6\Omega_{\varepsilon}}-1}{3\Omega_{\varepsilon}},~\,\&\,~\Omega_{\Lambda}{=\Omega_{k}}{=\Omega_{\hh}}{=0}\}\,.
\ea
\ee
That is to say,  in contrast to GR,  negative spatial curvature,  nontrivial magnetic $H$-flux,  negative cosmological constant, and  ordinary matter with sufficiently positive pressure   can all make    the expansion of the Universe accelerate.   

Besides, as seen from (\ref{decq2}),  the  time-evolution of the dilaton $({\dot{\phi}\neq0}$) also contributes to the acceleration, even though it needs to, and will be shown to, vary slowly.    In a nutshell,   the dilaton~$\phi$ has an opposite sign for its kinetic term in the string framed action (\ref{Reff}) and on-shell~(\ref{dotphiOmega}) this causes  the (desired) acceleration in   string frame.

If the time-varying of the dilaton is indeed small as of present, from  (\ref{dotphiOmega}),   the sum of the four density parameters should be close to unity,
 \be
 \Omega_{k}+\Omega_{\rhonew}+\Omega_{\Lambda}+\Omega_{\hh}\,\Big|_{\mathrm{at~}t=t_{0}}\simeq 1\,.
 \label{sumO}
 \ee
This allows us to  linearly approximate the deceleration parameter in (\ref{decq}) for the lower minus  sign and for  multi-component fluids as
\be
q_{0}\,\simeq\, 1-\Omega_{k}-5\Omega_{\hh}+\Omega_{\Lambda}+\sum_{w}\,(1{-3w})\Omega_{\rhonew, w}\,,
\label{q0}
\ee
which can also easily  assume a negative value, despite the  slow  variation of the dilaton. %\\ %\newpage

\subsection{Deceleration Naturally  in Einstein Frame}
Exclusively in this subsection,  to contrast the  physical distinction  of the frames, we postulate \textit{rather incorrectly} that a point particle would couple minimally  to the Einstein frame metric as in  the right hand side action of  (\ref{ptc2}).  The goal is to derive the Hubble and deceleration parameters in the Einstein frame, in comparison with those in the string frame.    

Firstly, from    $g^{\rm{E}}_{\mu\nu}=e^{-2\phi}g_{\mu\nu}$~(\ref{ggE}),  we note
\be
\ba{ll}
N_{\rm{E}}=e^{-\phi}N\,,\quad&\qquad a_{\rm{E}}=e^{-\phi}a\,,
\ea
\ee
and thus,
\be
H_{\rm{E}}=\frac{\,\rd \ln a_{\rm{E}}}{N_{\rm{E}}\rd t}=e^{\phi}\left(H-\frac{1}{N}\frac{\rd\phi}{\rd t}\right)\,.
\label{HEH}
\ee
Clearly, the Hubble parameter is not invariant under the change of the frames, $H_{\rm{E}}\neq H$,   
though it is a measurable  physical quantity.

We continue to compute the deceleration parameter in the Einstein frame,
\be
\ba{rll}
q_{\rm{E}}&\!=&\!\scalebox{1}{${-
\frac{1}{H_{\rm{E}}^{2}a_{\rm{E}}}}\left(\frac{\rmd~}{N_{\rm{E}\,}\rmd t}\right)^{2}a_{\rm{E}}$}\\
{}&\!=&\!\scalebox{1}{${{\frac{1}{6H^{2}_{\rm{E}}}\!\left[\left(\rhonew+3\pnew\right)e^{4\phi}+\frac{\hh^2}{a_{\rm{E}}^{6}}e^{-4\phi}+\left(\frac{2\rmd \phi}{N_{\rm{E}}\rmd t}\right)^{2}-\To e^{2\phi\,}\right]}}$}\,,
\ea
\label{AEq}
\ee
where the square of the Hubble parameter is determined through 
\be
H_{\rm{E}}^2=\frac{1}{3}\left(\frac{1}{N_{\rm{E}}}\frac{\rd\phi}{\rd t}\right)^2+\frac{1}{3}\rhonew e^{4\phi} +  \frac{1}{6}\To e^{2\phi} -\frac{k}{a_{\rm{E}}^{2}}+\frac{\hh^{2}}{12a_{\rm{E}}^{6}}e^{-4\phi}\,.
\label{HEsquared}
\ee
Like the Hubble parameter in (\ref{HEH}), contrasting  (\ref{AEq}) with  (\ref{decq}),   we note that  the physically measurable deceleration parameter is  frame-dependent too.    Not surprisingly,  the acceleration is absent and the deceleration is natural  in the Einstein frame: Unless  the strong energy condition $\rhonew+3\pnew>0$ is violated or the positive cosmological constant $\To=2\Lambda>0$ is dominant, $q_{\rm{E}}$ is generically  positive  (\textit{c.f.~}\cite{Marconnet:2022fmx,Capozziello:2006dj,Bahamonde:2017kbs}).

It is remarkable that, depending on the frame,  string or Einstein respectively,   the dilaton's  kinetic ``energy'' $\big(\frac{\rd\phi}{\rd t}\big)^2$,   the magnetic $H$-flux,  the negative cosmological constant ${\Lambda<0}$, and   positive pressure $\pnew>0$  accelerate or decelerate  the expansion of the Universe. Besides, while the  spatial curvature can change the sign of the deceleration parameter in the string frame~(\ref{decq2}), it cannot do so in the Einstein frame~(\ref{AEq}). It can merely affect the   value of the Hubble parameter squared~(\ref{HEsquared}).  %\\

%\newpage

\subsection{Differential Equations for the Redshift}
As   the observational data in cosmology  are typically parametrised by the redshift $z$ rather than time $t$, with $a=({1+z})^{-1}$ using 
\be
\frac{1}{N}\frac{\rd~}{\rd t}=-(1+z)H\,\frac{\rd~}{\rd z}\,,
\ee
it is useful to transform the $\ODD$-complete Friemann equations~(\ref{OFE}) to  differential equations of the redshift,
\be
\ba{rll}
\frac{\rd \phi(z)}{\rd z}&=&-\,\frac{1}{2(1+z)}\left[\,
3\pm\frac{\sqrt{\cF(z)}}{H(z)}\,\right]\,,\\
\frac{\rd H(z)}{\rd z}&=&-\,\frac{1}{1+z}\!\left[\,\frac{\pnew(z) e^{2\phi(z)}
{-2k(1+z)^{2}+\frac{1}{2}\hh^{2}(1+z)^{6}}}{H(z)}\,\pm\sqrt{\cF(z)}\,\right],
\ea
\label{ASMOFE}
\ee
where we  set a shorthand notation,
\be
\cF(z):=3 H(z)^{2}+2 \rhonew(z) e^{2\phi(z)} + 2\Lambda -6k(1+z)^{2}+\frac{1}{2}\hh^{2}(1+z)^{6}\,,
\ee
and from (\ref{rhonewep})
\be
\rhonew(z)=\rhonewo(1+z)^{3(1+w)}\,,
\qquad\qquad \pnew(z)=w\rhonewo(1+z)^{3(1+w)}\,.
\ee
%~\\
%We have solved (\ref{ASMOFE}) numerically, using   so-called the Runge-Kutta 4th Order method, \\
%\newpage

\section{Bayesian Inference from Observational Data}
We  solve the $\ODD$-complete Friedmann equations~(\ref{ASMOFE})  numerically, using   so-called the Runge-Kutta 4th Order method,  and    perform  the analyses of   Bayesian Inference  (BI)  with   the observational data from the supernovae and quasars~\cite{Scolnic:2021amr,Riess:2021jrx,King:2012id,Wilczynska:2015una,Martins:2017qxd,Wilczynska:2020rxx}.    We use the affine invariant Markov Chain Monte Carlo (MCMC) ensemble sampler called  \textsf{emcee}~\cite{Foreman-Mackey:2012any}. With  $100$ walkers, we run the samplers on a  supercomputer for $10^{6}$ steps to explore  maximal  six parameter space.  We discard $5\%$ of the initial steps as a ``burn-in'' phase, after which the MCMC chains are well converged. The autocorrelation time for each parameter is at most   $10370$ step---which is the case with FIG.\,\ref{FIGBI6pwfree}---hence sufficiently small. The final values  are determined to be the median ($50\%$) of the converged portion of the MCMC, while the  errors are denoted  at one-sigma level (from $16\%$ to $84\%$).  

\begin{figure}[H]
\!\!\!\includegraphics[width=1\linewidth]{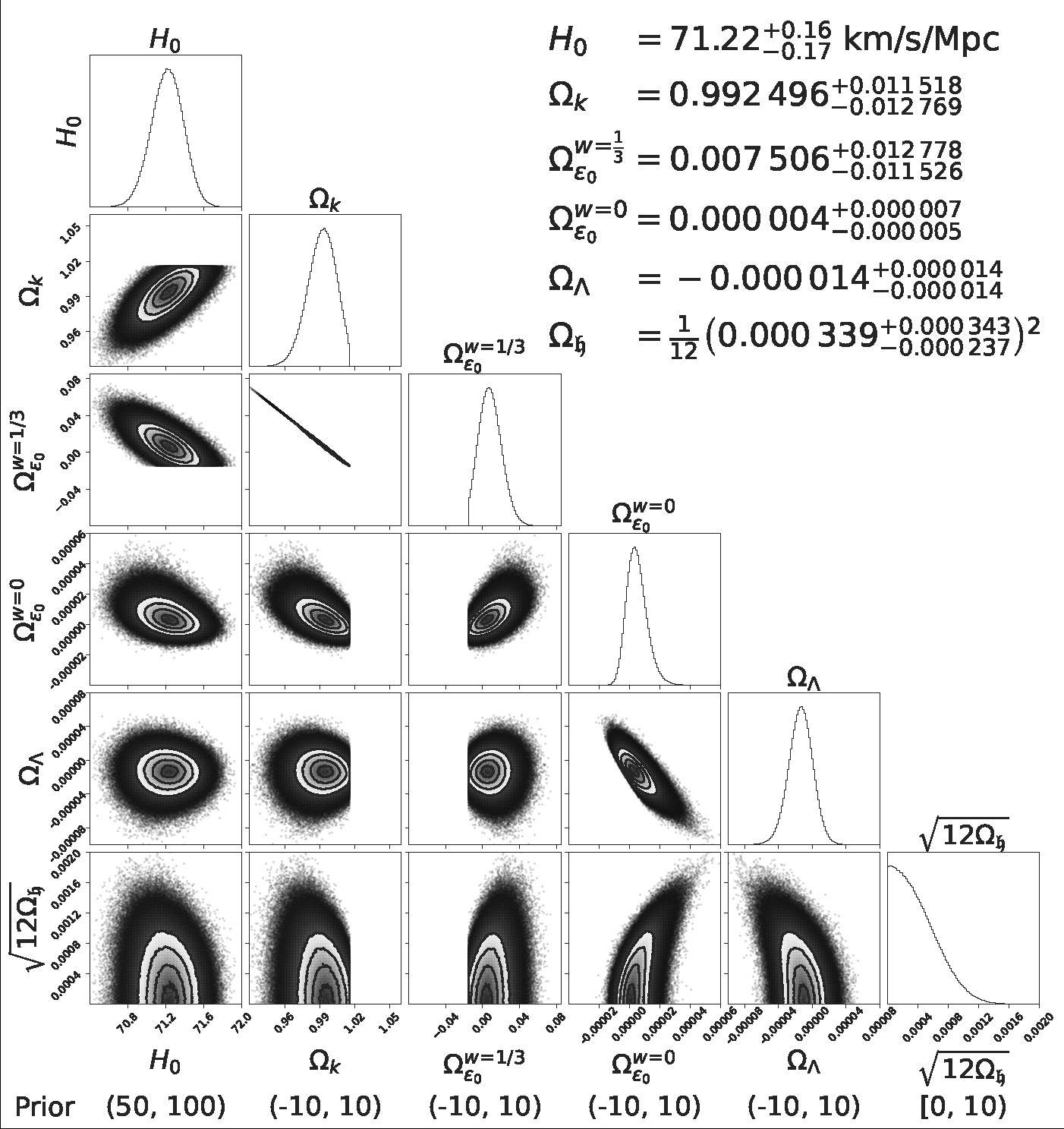}
\caption{{\bf{Bayesian Inference with  Six Parameters I:}} 
The priors are further cut for  the real valued  analysis of (\ref{OFE}) and FIG.\,\ref{FIGmu}.   }
\label{FIGBI6p}
\end{figure}

\begin{figure}[H]
\!\!\!\includegraphics[width=1\linewidth]{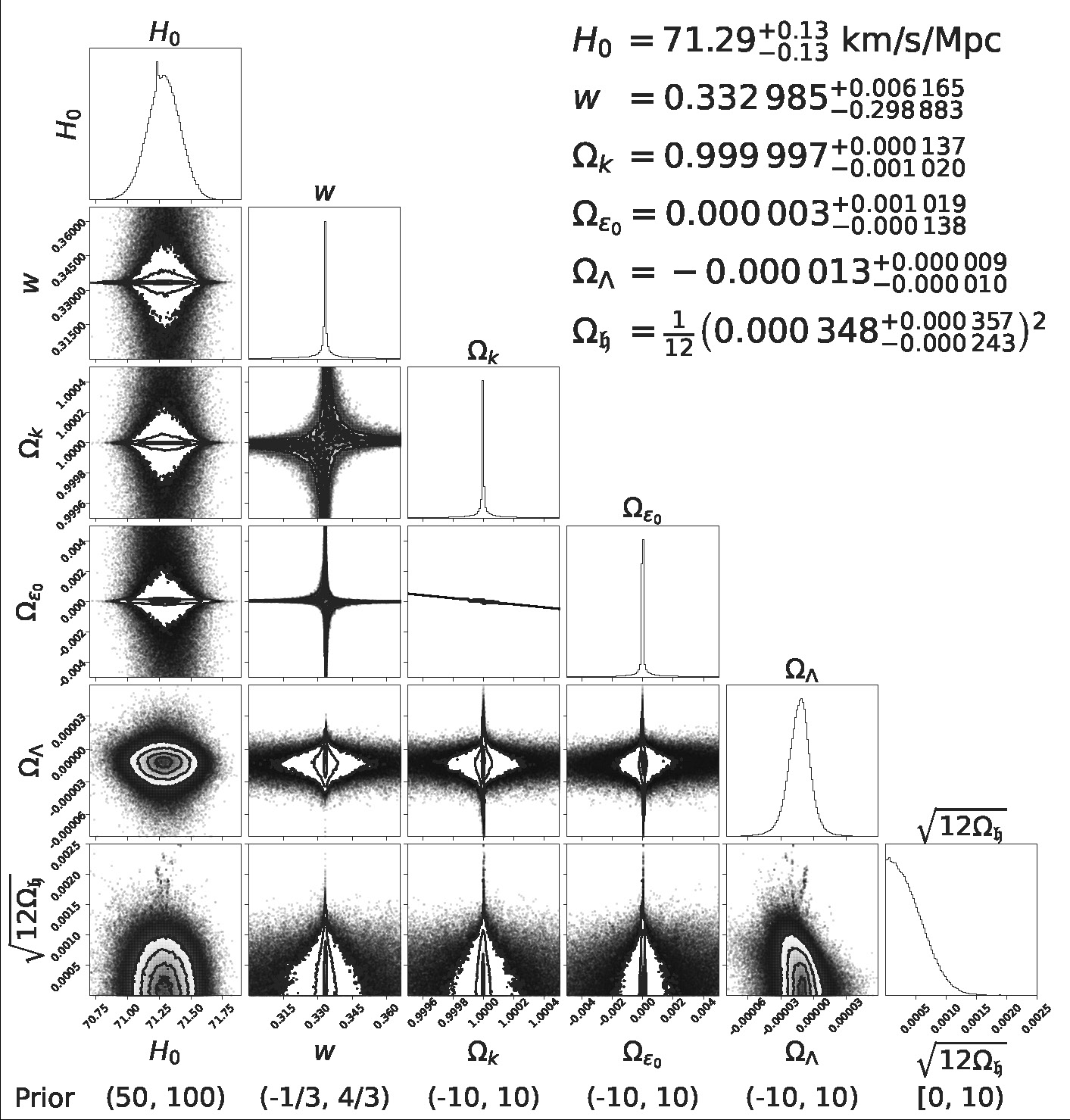}
\caption{{\bf{Bayesian Inference with  Six Parameters II:}}  
Up to one-sigma  error, FIG.\,\ref{FIGBI6p} and FIG.\,\ref{FIGBI6pwfree}  are consistent. We provide additional BI analyses in  the APPENDIX.}
\label{FIGBI6pwfree}
\end{figure}

 We  present our BI results  through  {FIGs.\,\ref{FIGBI6p},\,\ref{FIGBI6pwfree},\,\ref{FIGBI2p}}.  In sharp contrast to $\Lambda$CDM,  at present time   the curvature density    outstandingly  dominates, ${\Omega_{k}\approx 1}$ implying an open Universe,  while other  density parameters are all  negligible. We shall see later~(\ref{nocoin}) that these are not coincidental.  Subsequently,  from (\ref{ForOmega})  the sum of the density parameters at ${z=0}$  is   close to unity~(\ref{sumO}),   the  variation of the dilaton   $\dot{\phi}/N$  is indeed  suppressed~(\ref{dotphiOmega}), and the linear approximation of the   deceleration parameter (\ref{q0}) is valid. %\newpage

FIG.\,\ref{FIGBI6p}  illustrates our preliminary  BI analysis   where  we  postulate, with ${\To=2\Lambda}$, a pair of   perfect fluids, one with ${w=1/3}$ (radiation-like/pseudo-conformal) and the other with ${w=0}$ (pressureless). We have the  maximal  six  parameters, $\big\{H_{0},k,{\rhonewthird},\,{\rhonewzero},\,\Lambda,\hh\big\}$ at  ${z=0}$.

We  recognise     the relative dominance of the pseudo-conformal matter (${w=1/3}$)  over the pressureless dust~(${w=0}$) and   the  insignificance of the   cosmological constant, 
\be
{\Omega^{w=0}_{\rhonewo}<<\Omega^{w=1/3}_{\rhonewo}}\quad\mbox{~~and~~}\quad  \left| \Omega_{\Lambda}\right|<<\Omega^{w=1/3}_{\rhonewo}\,.
\ee 
Remarkably, these results  concur with     the  astrophysical  constraint~(\ref{SolarTest}) that  is the condition for DFT to pass the solar system test, since  from  $K_{\mu}{}^{\mu}=(1{-3w})K_{t}{}^{t}$ and  ${\To= 2\Lambda}$,  the following relation holds
\be
\frac{K_{\mu}{}^{\mu}-\To}{-K_{t}{}^{t}}=3w-1-2\frac{\Omega_{\Lambda}}{\Omega_{\varepsilon}}\,.
\ee
According to the BI  results of FIG.\,\ref{FIGBI6p},  for the dominant pseudo-conformal matter,  this has the median  $4\times 10^{-3}$ and further can be  zero at the one sigma level.\newpage

\newpage

We emphasize that our inference for an open Universe \textit{i.e.~}$\Omega_{k}\approx 1$, as opposed to a flat or closed one, is unbiased. This is because we set the prior support for $\Omega_{k}$ to range from negative to positive values, specifically from $-10$ to $+10$.  On theoretical grounds, an open Universe tends to accelerate the expansion of Universe, as we illustrated  in  section~\ref{SECODD},  and furthermore causes the time-varying dilaton to slow down to a constant, as we discuss in the following  section~\ref{ExactV}.  

The median   $\Omega^{w=1/3}_{\rhonewo}\simeq 0.007$ coincides with  some known luminous baryonic matter  density~\cite{FukugitaPeebles}, which further  allows   little room for dark matter. 
 
To  confirm the  bias toward  ${w={1}/{3}}$ against ${w=0}$,   we perform a  complementary BI analysis with different six parameters, $\big\{H_{0},k,w,{\rhonewo},\Lambda,\hh\big\}$, now  focusing on  a single   fluid having  an arbitrary equation-of-state parameter $w$.  The result is depicted in FIG.\,\ref{FIGBI6pwfree}. Though the uncertainty is relatively higher, it   shows persistently  $w\simeq 1/3$.\\~\\% \newpage

\section{Exact Vacuum  Free of  Coincidence Problem\label{ExactV}}  
The  above Bayesian Inference  results   indicate  that ${\Omega_{\rhonewo}}$'s  and ${\Omega_{\Lambda}}$ are all  highly suppressed. This motivates  us to
closely examine the vacuum geometry of stringy gravity, which is free of matter and the cosmological constant but has ${k < 0}$. Fortunately, an analytic solution  is available in the literature \cite{Copeland:1994vi} in terms of conformal time $\eta$.  Here, we further elaborate on this known solution to feature three free parameters  $\big\{H_{0},\,\hh,\,l\equiv 1/\sqrt{-k}\,\big\}$, or $\big\{H_{0},\,\Omega_{\hh},\,\Omega_{k}\,\big\}$, in addition to the redundant parameters $a_{0}$ and $\phi_{0}$.  We refer to the APPENDIX for our own  derivation.  We simply   spell the solution below. 

 The   vacuum geometry  is characterised by  the triplet: the dilaton  that does \textit{not}  run away because $\,k<0$,
\be
\ba{l}
e^{2\phi(\eta)-2\phi_{0}}=\frac{1}{2}
\left(1{+\sigma}\sqrt{1-\frac{(\hh l\sinh\zeta)^2}{12a_{0}^{4}}}\right)\!\left(
\frac{\tanh\left(\frac{\eta-\eta_{0}}{l}+\frac{\zeta}{2}\right)}{\tanh\frac{\zeta}{2}}\right)^{\!\sqrt{3}}\vspace{7pt}\\
\qquad~+\frac{1}{2}
\left(1{-\sigma}\sqrt{1-\frac{(\hh l\sinh\zeta)^2}{12a_{0}^{4}}}\right)\!\left(
\frac{\tanh\left(\frac{\eta-\eta_{0}}{l}+\frac{\zeta}{2}\right)}{\tanh\frac{\zeta}{2}}\right)^{\!-\sqrt{3}\,}\,,
\ea
\label{analyticphi}
\ee
the (homogeneous \& isotropic)   magnetic $H$-flux,
\be
H_{(3)}=\frac{\hh\, r^2 \sin\vartheta\,\rd r\wedge\rd\vartheta\wedge\rd\varphi}{\left[1-(r/2l)^2\right]^3}\,,
\label{mHflux}
\ee
and the Friedmann--Lema\^{i}tre--Robertson--Walker metric,
\be
\rd s^2=a^2(\eta)\left[-\rd \eta^2+\frac{\rd r^2+r^2\rd\vartheta^2+r^2{\sin^2\!\vartheta}\rd\varphi^2}{\left\{1-(r/2l)^2\right\}^2}\right]\,,
\label{FRWconformal}
\ee
having the  scale factor, 
\be
a^{2}(\eta)=a_{0}^{2\,}e^{2\phi(\eta)-2\phi_{0}\,}\frac{\sinh\left({2(\eta-\eta_{0})}/{l}+\zeta\right)}{\sinh\zeta}\,.
\label{SF}
\ee
Here the coordinate chart is isotropic~(\ref{AIconversion}) and conformal,  $\zeta$ is  a constant, and  $\sigma$ is another  sign factor   which  can be  all fixed   by  $\Omega_{k}$,  
$\Omega_{\hh}$ at $\eta=\eta_{0}$,  and  the $\pm$ sign in (\ref{OFE}): From the expressions  of the Hubble constant,
\be
H_{0}=\frac{1}{2a_{0}l\sinh\zeta}\left[2\cosh\zeta+
{\sigma\sqrt{12-a_{0}^{-4}\left(\hh l\sinh\zeta\right)^2}}
\right]\,,
\ee
and the two density parameters,
\be
\ba{ll}
\dis{\left.\Omega_{k}\right|_{\mathrm{\,at~}\eta=\eta_{0}}=\frac{1}{l^2a_{0}^2H_{0}^2}\,,}\qquad&\qquad
\dis{\left.\Omega_{\hh}\right|_{\mathrm{\,at~}\eta=\eta_{0}}=\frac{\hh^2}{12a_{0}^6H_{0}^2}\,,}
\ea
\ee
one can derive 
\be
\ba{rll}
\sinh\zeta&=&
\left.\mp\sqrt{\frac{2\Omega_{k}}{2+\Omega_{k}+3\Omega_{\hh}\pm\sqrt{3+6\Omega_{k}+6\Omega_{\hh}}}}~\right|_{\mathrm{\,at~}\eta=\eta_{0}}\,,\\
\sigma&=&\left.-\,\frac{\sqrt{1+2\Omega_{k}+2\Omega_{\hh}}\pm\sqrt{3}\,}{\,\left|\sqrt{1+2\Omega_{k}+2\Omega_{\hh}}\pm\sqrt{3}\,\right|}~
\right|_{\mathrm{\,at~}\eta=\eta_{0}}\,.
\ea
\label{szs}
\ee
Thus,   for  the upper sign we always have $\sigma =-1$.   For the lower  sign (of our interest), we  set  ${\sigma=+1}$ if  $\,\Omega_{k}+\Omega_{\hh}< 1$; Otherwise    $\sigma=-1$.
\begin{comment}
Conversely, the  two density parameters  and the Hubble constant at present time are related to  $\big\{l,\hh,\zeta,a_{0},\sigma\big\}$,
\be
\ba{ll}\vspace{3pt}
\dis{\left.\Omega_{k}\right|_{\mathrm{\,at~}\eta=\eta_{0}}=\frac{1}{l^2a_{0}^2H_{0}^2}\,,}\qquad&\qquad
\dis{\left.\Omega_{\hh}\right|_{\mathrm{\,at~}\eta=\eta_{0}}=\frac{\hh^2}{12a_{0}^6H_{0}^2}\,,}\\
\multicolumn{2}{c}{\dis{
H_{0}=\frac{1}{2a_{0}l\sinh\zeta}\left[2\cosh\zeta+
{\sigma\sqrt{12-a_{0}^{-4}\left(\hh l\sinh\zeta\right)^2}}\right]\,.}}
\ea
\ee
\end{comment}
    
It is worth while to note that,  replacing $l$ and $\zeta$ by imaginary numbers $-il$ and $i\zeta$, one can obtain the exact geometry of a closed Universe $k>0$ (see \cite{Copeland:1994vi} for the explicit expression).  However,  this will convert the converging hyperbolic tangent functions in (\ref{analyticphi}) to diverging  tangent functions, such that the dilaton $\phi$ will not be stabilised over the cosmic evolution, which conflicts with the observational constraint~(\ref{alphaphi}). When ${k=0}$, the dilaton is linear in time which is  unacceptable too. It is essential that the Universe should be open in order to make the dilaton~$\phi$ slowly varying and convergent.\\

\subsection{Bayesian Inference with Exact Vacuum}
\begin{figure}[H]
	\centering
	\includegraphics[width=0.93\linewidth]{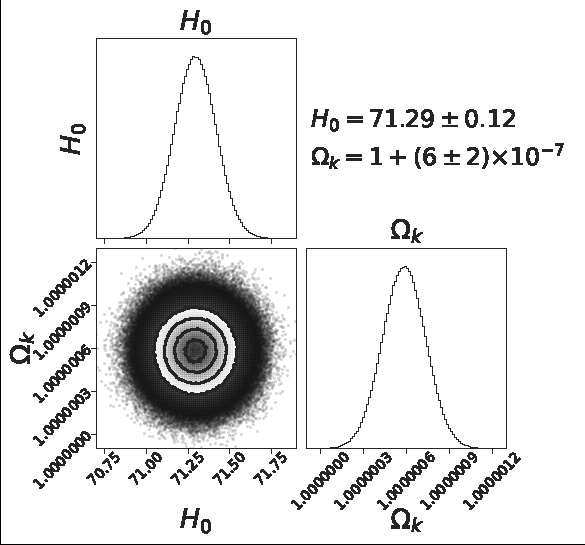}\vspace{-5pt}
\caption{{\bf{Bayesian Inference with Two  Parameters:}}  \\
Minimal Vacuum Geometry with ${\hh\equiv0}$, ${q_{0}<0}$, ${\sigma=-1}$.} 
\label{FIGBI2p}
\end{figure}

FIG.\,\ref{FIGBI2p} represents our well converged, minimal  BI analysis for the $H$-flux-free vacuum geometry with  only  two  parameters, $H_{0}$ and $\Omega_{k}$.  The  vacuum is clearly not a   de Sitter space which is a natural solution in GR but not in DFT, \textit{c.f.~}\cite{Danielsson:2018ztv,Obied:2018sgi,Agrawal:2018own,Andriot:2018wzk,Garg:2018reu,Murayama:2018lie,Hamaguchi:2018vtv,Ooguri:2018wrx,Hebecker:2018vxz,Bedroya:2019tba,Bedroya:2020rac,Bena:2023sks}. % which supports recent    `de Sitter  swampland'    conjecture~\cite{Danielsson:2018ztv,Obied:2018sgi,Agrawal:2018own,Andriot:2018wzk,Garg:2018reu,Murayama:2018lie,Hamaguchi:2018vtv,Ooguri:2018wrx,Hebecker:2018vxz,Bedroya:2019tba,Bedroya:2020rac,Bena:2023sks}.  
~\\
~\\

\subsection{Prediction for Future Infinity}
From the three  BI results  of  {FIGs.\,\ref{FIGBI6p},\,\ref{FIGBI6pwfree},\,\ref{FIGBI2p}},  we deduce that,  at present the  Universe enters a vacuum-dominated era  with the   two significant  parameters $\{H_{0},\Omega_{k}\}$. Then,  resorting to the  analytic  solution, we can  predict that at   future infinity, \textit{i)} the dilaton converges to  a constant,
\be
\dis{\lim_{\eta\rightarrow\infty}}e^{2\phi(\eta)-2\phi_{0}}
=\textstyle{
\frac{1{+\sigma}\sqrt{1-\frac{(\hh l\sinh\zeta)^2}{12a_{0}^{4}}}}{2\tanh^{\sqrt{3}}(\zeta/2)}+\frac{
1{-\sigma}\sqrt{1-\frac{(\hh l\sinh\zeta)^2}{12a_{0}^{4}}}}{2\tanh^{-\sqrt{3}}(\zeta/2)}}\,;
\label{limitphi}
\ee
\textit{ii)} the  Universe expands forever as 
 \be
 {N(\eta)=a(\eta)\propto e^{\eta/l}}\qquad\mbox{for~~large~~}\eta\,;
 \ee 
\textit{iii)} the Hubble and  deceleration parameters both vanish,%\footnote{See APPENDIX for our illustrating figures.}
\be
\ba{lll}
\dis{\lim_{\eta\rightarrow\infty} H(\eta)=0}\,,~&~
\dis{\lim_{\eta\rightarrow\infty} q(\eta)=0}\,,~&~
\dis{\lim_{\eta\rightarrow\infty} a(\eta)H(\eta)=\frac{1}{l}}\,;
\ea
\label{FutureInfty}
\ee
 and  \textit{iv)} with ${\Lambda\equiv 0}$ the density parameters converge to either unity or zero,
\be
\lim_{\eta\rightarrow\infty}\Omega_{k}=1\,,\qquad\quad
\lim_{\eta\rightarrow\infty}\Omega_{\mathbf{others}}=0\,.
\label{nocoin}
\ee
The density parameter $\Omega_{\Lambda}$ would diverge in the limit unless $\Lambda=0$.  The cosmological constant should  better be zero. The limiting behaviors of all the density parameters are then all  consistent with  our BI fitting  and show that there is no coincidence problem in our approach.

The converging value for the dilaton~(\ref{limitphi})  corresponds to the final value of the fine  structure constant at future infinity normalised by the  present value since $\alpha(\eta)/\alpha(\eta_{0})=e^{2\phi(\eta)-2\phi_{0}}$~(\ref{alphaphi}). \\

%\newpage

%%%%%%%%%%%%%%%%%%%%%%%%%%%%%%
%%%%%%%%%%%%%%%%%%%%%%%%%%%%%%
%%%%%%%%%%%%%%%%%%%%%%%%%%%%%%
%\vspace{3pt}

\section{Confronting with the Observations}  
The  light  trajectory  in the open Universe of $k=-1/l^{2}$  satisfies, with (\ref{Arrr}),
\be
\ba{l}
\dis{\int^{z}_{0}}\frac{\rd z^{\prime}}{H(z^{\prime})}=
\dis{\int_{0}^{\rr}}\frac{\rd\rr^{\prime}}{\sqrt{1+\rr^{\prime 2}/l^2}}
=l\sinh^{-1}(\rr/l)\\
=l\ln\left(
{\rr}/{l}+\sqrt{1+{\rr^2}/{l^2}}\right)=2l\ln\left[\frac{1+(r/2l)}{\sqrt{1-(r/2l)^2}}\right]\,,
\ea
\ee
and thus, the luminosity distance is given by
\be
\ba{rll}d_{L}(z)&=&\dis{\frac{\rr}{a(z)}\,=\,\frac{r}{a(z)\big[1-(r/2l)^{2}\big]}}\\
{}&=&\dis{l(1+z)\sinh\!\left[\dis{\frac{1}{l}\int^{z}_{0}}\!\frac{\rd z^{\prime}}{H(z^{\prime})}\right]}\,,
\ea
\ee
which gives the distance modulus,
\be
\mu(z)=5\,\mbox{Log}_{10}\left[\,\frac{d_{L}(z)}{10\,\rm{pc}}\,\right]\,.
\label{distancemodulus}
\ee

\begin{figure}[H]
	\centering
	\includegraphics[width=1\linewidth]{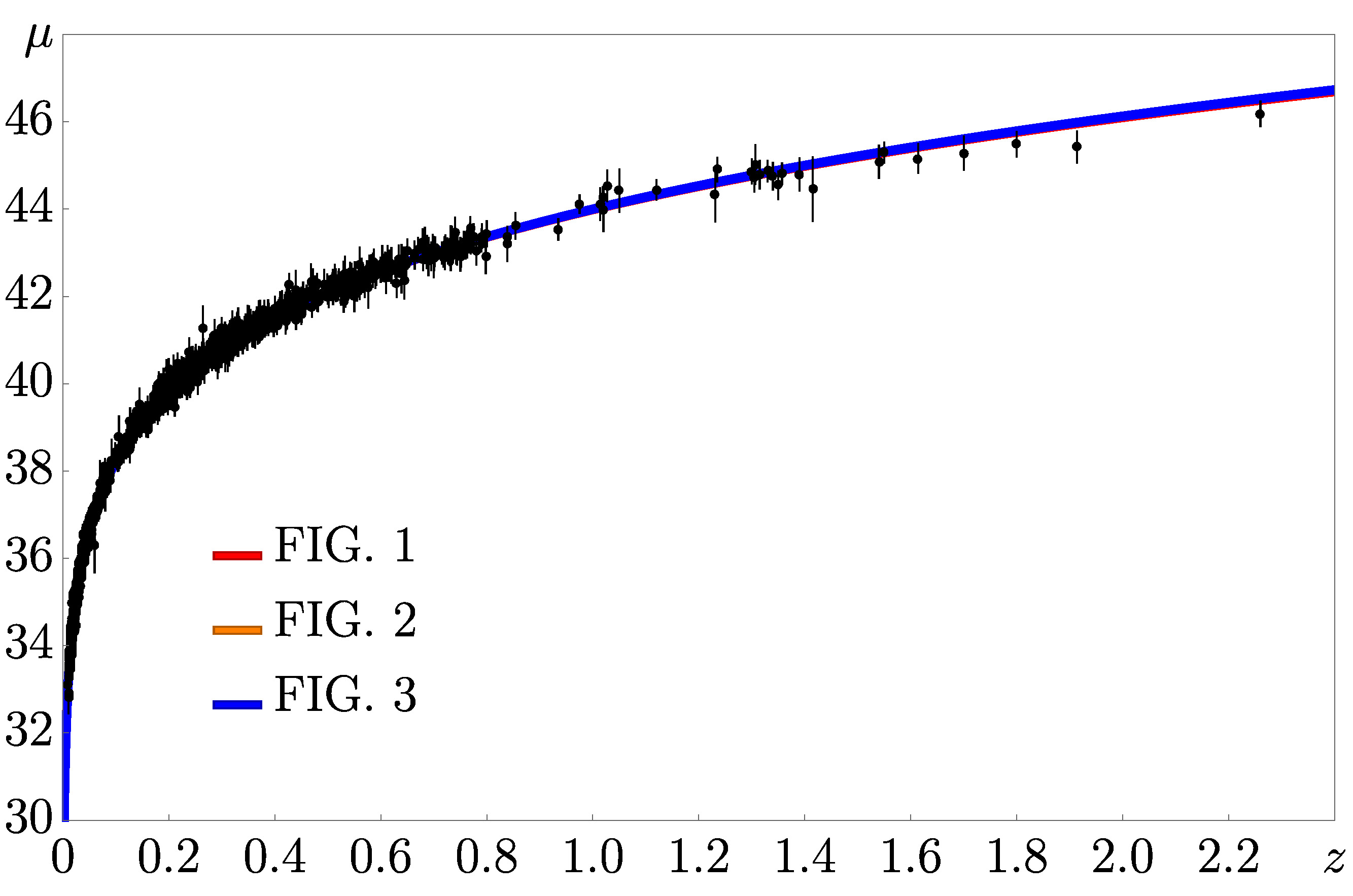}
\caption{{\bf{Distance Modulus:}} Complete agreement of all the three BI results in   FIGs.\,\ref{FIGBI6p},\,\ref{FIGBI6pwfree},\,\ref{FIGBI2p}  with the  SNe Ia data~\cite{Scolnic:2021amr,Riess:2021jrx}.}
	\label{FIGmu}
	\end{figure}

\begin{figure}[H]
	\centering
	\includegraphics[width=1\linewidth]{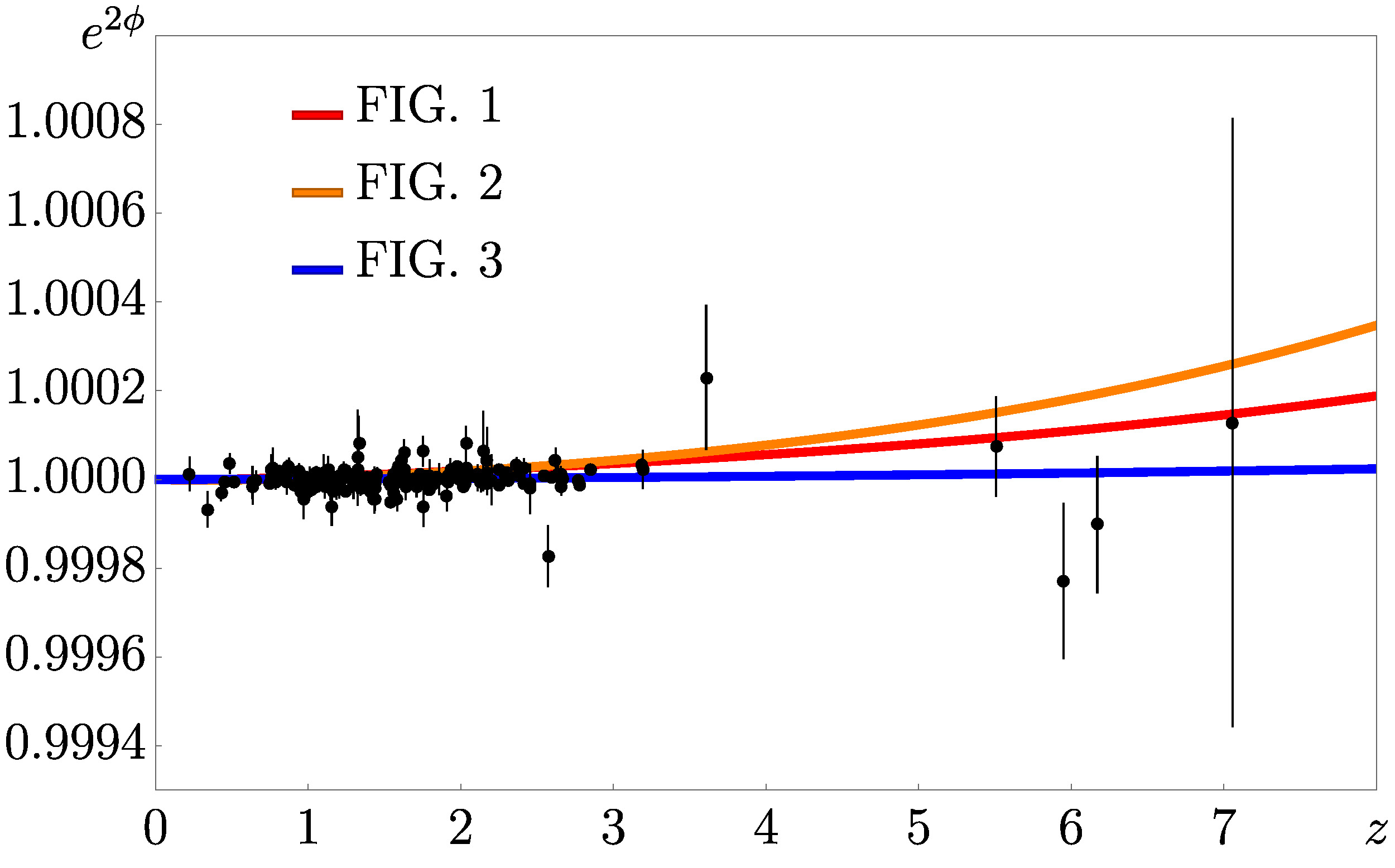}
\caption{{\bf{Fine-structure constant: }}Consistency of the three BI results in FIGs.\,\ref{FIGBI6p},\,\ref{FIGBI6pwfree},\,\ref{FIGBI2p}  with  the quasar data~\cite{King:2012id,Wilczynska:2015una,Martins:2017qxd,Wilczynska:2020rxx}. }
	\label{FIGphiz}
	\end{figure}

We now confront the BI results shown in FIGs. \ref{FIGBI6p}, \ref{FIGBI6pwfree}, and \ref{FIGBI2p} with the observed distance modulus $\mu(z)$~\cite{Scolnic:2021amr, Riess:2021jrx} and the evolution of the fine-structure constant~(\ref{alphaphi})~\cite{King:2012id, Wilczynska:2015una, Martins:2017qxd, Wilczynska:2020rxx} in FIGs. \ref{FIGmu} and \ref{FIGphiz}, respectively. We find excellent agreement.


\subsection*{Negligible Lorentz Symmetry Breaking}
The magnetic $H$-flux~(\ref{mHflux})  implies \textit{a priori} a Lorentz symmetry breaking term in (\ref{QED}): at  ${r=0}$ at present,
\be
\frac{1}{24}H_{\mu\nu\rho}\bar{\psi}\gamma^{\mu\nu\rho}\psi=\frac{1}{4}H_{123}\bar{\psi}\gamma^{123}\psi=
i\frac{1}{4}\hh\,\bar{\psi}\gamma^{5}\gamma^{0}\psi\,.  
\ee
The strength of the $H$-flux we have estimated in FIG.\,\ref{FIGBI6p}  is  safely    far less than the stringent  experimental  bound~\cite{Ferrari:2018tps} (our $\hh/4$ is ``$b_{0}$'' therein, see also \cite{Roberts:2014dda,Kostelecky:2008ts}):
\be
|\hh|\,\simeq\, 3\times 10^{-4} H_{0}\,\simeq\, 2{\times 10^{-36}}\,\mathrm{eV} ~\,{<<}\,~1.5{\times 10^{-11}}\,\mathrm{eV}\,.
\ee
~\\

\section{Conclusion} 
To summarise, the gravitational theory of the closed string massless sector agrees admirably  well with the late-time  ($z\lesssim 7$) cosmological data~\cite{Scolnic:2021amr,Riess:2021jrx,King:2012id,Wilczynska:2015una,Martins:2017qxd,Wilczynska:2020rxx}, without any need for dark energy or dark matter, free of any coincidence~(\ref{nocoin}). The only requirements are an open Universe and a string frame.  We stress that the dilaton  started to vary slow converging  eventually to a constant~(\ref{limitphi}) in the late-time Universe, because the Universe is open (${k=-1/l^2<0}$), which is clear from the analytic solution~(\ref{analyticphi}).  We estimate   the Hubble constant as  $H_{0}\simeq 71.2\pm 0.2\,\mathrm{km/s/Mpc}$,  and the  spatial curvature length scale  as $l=1/\sqrt{-k}\,\simeq 1/H_{0}\simeq 4.2\,\mathrm{Gpc}$.  We also assert that the cosmological constant ought to be  exactly zero, in order to ensure a coincidence-free cosmology.

\begin{figure}[h]
	\centering
	\includegraphics[width=1\linewidth]{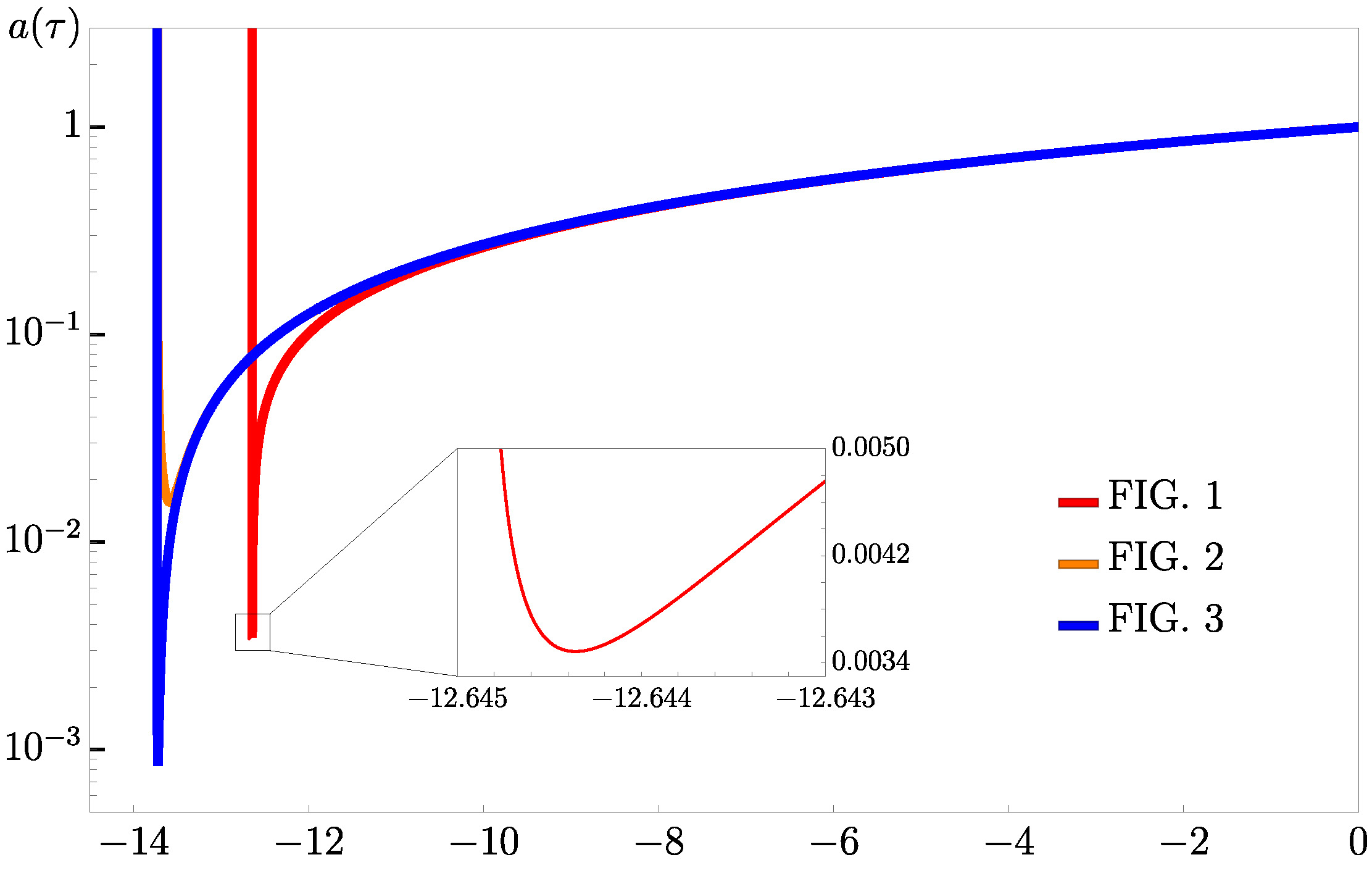}\vspace{-5pt}
\caption{{\bf{Scale Factor of the Bouncing  Open Universe.}} The bouncing occurs   $12.64$, $13.58$, or $  13.72$ gigayears ago (cosmic time~$\tau$) for   FIGs.\,\ref{FIGBI6p},\,\ref{FIGBI6pwfree},\,\ref{FIGBI2p} respectively.   Coincidentally,  these are comparable to    the ``age" of the flat Universe  in $\Lambda$CDM.  } 
\label{FIGa}
\end{figure}
Whenever the Hubble parameter vanishes, with  $\pnew\geq 0$,  Eq.(\ref{OFE}) implies a \textit{bouncing} open Universe   (\textit{cf.~}\cite{Brandenberger:2016vhg,Brandenberger:2023wtd,Brandenberger:2023ver}):
\be
\left.{\frac{1}{N}\frac{\rd H}{\rd t}}\right|_{{\rm{when\,}}H=0}=\pnew e^{2\phi}
+\frac{2}{l^2a^{2}}+\frac{\hh^{2}}{2a^{6}}\,>\,0\,,
\label{bouncing}
\ee
featuring  inflation-like acceleration,~${q \rightarrow -\infty}$. This is indeed the case with   all our BI analyses elongated to higher redshifts, including the vacuum geometry~(\ref{SF}), 
see FIG.\,\ref{FIGa}. The big bang singularity in $\Lambda$CDM is   replaced by a smooth bouncing around $13$ gigayears ago, without resort to any  quantum or stringy     $\alpha^{\prime}$  correction (\textit{c.f.~e.g.}~\cite{Hohm:2019ccp,Hohm:2019jgu,Bernardo:2019bkz,Bernardo:2020zlc,Bernardo:2021xtr,Quintin:2021eup,Lescano:2022nrb}). The bouncing occurs only in the string frame where the kinetic term of the dilaton has the opposite sign  to generate the required acceleration, as illustrated in  section~\ref{SECODD},

Although the deceleration parameter can be easily negative  in theory~(\ref{q0}),  our BI fitting results indicate that the present time acceleration is  marginal: $q_{0}$ amounts to $0.0075\pm 0.0128$,   $- 10^{-5}$, or  $- 6{\times 10^{-7}}$ for FIGs.\,\ref{FIGBI6p}, \ref{FIGBI6pwfree}, and \ref{FIGBI2p} respectively.  This, along with the BI result of $\Omega_{k}\simeq 1$ (hence $a_{0}H_{0}\simeq 1/l$), appears  to imply that the present Universe is already in the converging stage of (\ref{FutureInfty}), and therefore there is no issue of coincidence.

% Nevertheless,  $q \rightarrow -\infty$  when  bouncing~(\ref{bouncing}).

Our work is to be a crucial step in validating the gravity of  string theory~(\ref{EDFEODD}), a special kind of modified gravity that comes from a top-down approach based on the $\ODD$ symmetry principle. However, this work is not definitive and requires further corroboration from diverse data sources, including the cosmic microwave background of the early Universe~\cite{Planck:2018vyg}. The theory's ability to explain the earlier Universe remains an open question for further investigation.

{It is a significant outcome of our BI analysis that a radiation-like substance, characterized by an equation-of-state parameter ${w =1/3}$, is more prevalent than  pressureless dust which has ${w = 0}$. However, the cumulative density of all matter is trivial as $\Omega_{\varepsilon} \ll \Omega_{k}$. This stands in sharp contrast to the conventional  understanding of ordinary matter or cold dark matter.  Perhaps, the most noteworthy result of this paper is  that the vacuum geometry of the string theory gravity  aligns remarkably well with the observed late-time cosmological data.   We must acknowledge a limitation in our BI analysis with matter, while drawing attention to a recent nuclear physics experiment~\cite{Burkert:2018bqq} in the remaining of this paper.

As has been stated below (\ref{defO}), for simplicity,  we initially assumed that only the cosmological constant contributes to $\To$,  without considering any matter contributions~(\ref{ToLF2}).  To improve upon this, we could include genuine  matter contributions  by introducing an additional equation-of-state parameter $\lambda=e^{-2\phi}\To/\varepsilon$ alongside $w ={\pnew}/{\varepsilon}$.  It is encouraging  that the continuity equation~(\ref{conservation}) can be still  resolved:
\be
\varepsilon(t) = \varepsilon_{0}a^{-\frac{6(w+1)}{\lambda+2}}e^{- \frac{2\lambda}{\lambda+2}\phi}.
\ee
In subsequent BI research, we intend to ascertain new values for $\lambda$ and $w$  that should more accurately represent the dominant matter.

On a different note, the experiment~\cite{Burkert:2018bqq} has demonstrated  that the core of a proton withstands pressures exceeding those in any other known form of matter, nearly ten times the pressure within a neutron star's core.   Cold baryons may exhibit negligible thermal pressure,  exerted by the low kinetic energy of their center of mass motion. Nevertheless, if the hadronic internal pressure is comparable to the energy density, as indicated by the experiment~\cite{Burkert:2018bqq}, then both the energy and pressure localized within hadrons may  source  the gravity.     While the experiment above deals with the  pressure and the energy density inside the hadron, it may have a repercussion on the values of the mass density $\varepsilon(t)$,  its spatial counterpart $\pnew(t)$,  and  the scalar component of the energy-momentum tensor $\To(t)$. If this is the case, for baryonic matter we may well write $\varepsilon(t)=n(t)m_{\varepsilon}$, $\pnew(t)=n(t)m_{\pnew}$, and $e^{-2\phi(t)}\To(t)=n(t)m_{\To}$,  where $n(t)$ is a (averaged) number density  and, from (\ref{rhonewpnew}),
\be
\ba{cll}
m_{\varepsilon}&=&\dis{-\int_{\rm{Baryon}}\!\!\!\!\!\!\!\!\!\!\!\!\rmd^3x~\sqrt{-g}e^{-2\phi}K_{t}{}^{t}}\,,\\
m_{\pnew}&=&\dis{\frac{1}{3}\int_{\rm{Baryon}}\!\!\!\!\!\!\!\!\!\!\!\!\rmd^3x~\sqrt{-g}e^{-2\phi}\left(\sum_{i=1}^{3}K_{i}{}^{i}\right)}\,,\\
m_{\To}&=&\dis{\int_{\rm{Baryon}}\!\!\!\!\!\!\!\!\!\!\!\!\rmd^3x~\sqrt{-g}e^{-2\phi}\To}\,.
\ea
\ee
The volume integral domain  is to be  over the interior of a baryon.  In particular,  $m_{\varepsilon}$ is the Newtonian mass of a single baryon~(\ref{NewtonMass}).  Note $w=m_{\pnew}/m_{\varepsilon}$ and $\lambda=m_{\To}/m_{\varepsilon}$ which are dimensionless and naturally constant. Further investigation is undoubtedly necessary.\\
}\\

\newpage

\noindent{\textit{Acknowledgments.}}---We  thank   Robert Brandenberger, Hun Jang,  Chung-Chi Lee, Hyun Kyu Lee,  Su Houng Lee,   Mannque Rho, and Stefano Scopel for  valuable  discussions at  various stages of the  project. 
We are  grateful  to Daniel Scolnic, Dillon Brout, and Anthony Carr for clarification regarding the Pantheon$\mathbf{+}$ sample.  The Bayesian analysis  was done through   KiSTi  National Supercomputing Center with kind  help from   Sunghyun Kang, In-Ho Lee, and  Kawon Lee.   We thank also Shehu Abdus Salam for discussions  regarding the analysis. This  work is supported by Basic Science Research Program through the National Research Foundation of Korea (NRF)  Grants, NRF-2023R1A2C2005360,  NRF-2023K2A9A1A01098740,   NRF-2020R1A6A1A03047877   (Center for Quantum Spacetime: CQUeST).  LY is  sponsored  by Young Scientist Training  Program  from Asia  Pacific Center for Theoretical Physics (APCTP) and further by  Natural Science Foundation of Shanghai 24ZR1424600.  Part of this work was performed at the workshop, \href{https://www.grp2023.apctpstring.com/}{`Gravity beyond Riemannian Paradigm'}, Jeju, June 2023,  coorganised by CQUeST and APCTP.\\
\hfill 
~\\
~\\

%\newpage

\appendix
\renewcommand{\arraystretch}{2} 
\renewcommand\thesection{A}
\renewcommand{\thefigure}{A\,\arabic{figure}}
\setlength{\jot}{9pt}                 % spacing btwn the rows of an eqnarray
\renewcommand{\arraystretch}{3.2} 
\setcounter{equation}{0}
%\begin{widetext}
\section*{--- APPENDIX ---}
\vspace{12pt}

\subsection*{Deriving the Analytic  Vacuum Geometry}
In the absence of matter  and the cosmological constant,  the stringy Friedmann equations~(\ref{OFE}) reduce to
\be
\ba{ll}
\frac{2\phi^{\prime}}{N}{-3H}=\pm\sqrt{3H^2{+
\frac{\hh^{2}}{2a^{6}}}{-\frac{6k}{a^{2}}}}\,,\quad&\quad
\frac{1}{N}\!\left(\frac{2\phi^{\prime}}{N}{-3H}\right)^{\prime}=3H^2
\,.
\ea
\label{ASMreduced2}
\ee
We focus on an open Universe with  $l=1/\sqrt{-k}>0$,  choose a conformal gauge with conformal time $\eta$, and set~\cite{Copeland:1994vi},
\be
N(\eta)\equiv a(\eta)\,,\qquad\qquad b(\eta)\equiv e^{-\phi(\eta)}a(\eta)\,.
\ee
In fact,  $b(\eta)$ is  the scale factor in the Einstein frame.  In terms of $\phi$ and $b$, the Hubble parameter is given by
\be
H=\left(\frac{\phi^{\prime}}{b}+\frac{b^{\prime}}{b^2}\right)e^{-\phi}\,,
\ee
and (\ref{ASMreduced2}) is equivalent to
\be
\ba{rll}
\phi^{\prime}+3\frac{b^{\prime}}{b\,}&=&\mp\sqrt{3\left(\phi^{\prime}+\frac{b^{\prime}}{b\,}\right)^2+\frac{6}{l^2}+\frac{\hh^{2}}{2b^4}e^{-4\phi}}\,,\\
\left[\frac{e^{-\phi}}{b}\left(\phi^{\prime}+3\frac{b^{\prime}}{b\,}\right)\right]^{\prime}&=&-3\frac{e^{-\phi}}{b}\left(\phi^{\prime}+\frac{b^{\prime}}{b\,}\right)^{2}\,.
\ea
\label{ASMtwob}
\ee
On one hand, the former of these equations means
\be
\phi^{\prime}{}^2=3\left(\frac{b^{\prime}}{b}\right)^2-\frac{3}{l^{2}}-\frac{\hh^{2}}{4b^4}e^{-4\phi}\,,
\label{ASMbeforeh}
\ee
and taking a derivative of this, we get
\be
\phi^{\prime}\phi^{\prime\prime}=3\left(\frac{b^{\prime}}{b}\right)\left(\frac{b^{\prime}}{b}\right)^{\prime}+\left(\phi^{\prime}+\frac{b^{\prime}}{b\,}\right)\frac{\hh^{2}}{2b^4}e^{-4\phi}\,.
\label{ASMafterh}
\ee

On the other hand, the latter of (\ref{ASMtwob})  gives
\be
\phi^{\prime\prime}+3\left(\frac{b^{\prime}}{b}\right)^{\prime}+2\phi^{\prime 2}+2\phi^{\prime}\frac{b^{\prime}}{b}=0\,.
\label{ASMphipp}
\ee
Substituting the expressions of $\frac{\hh^{2}}{4b^4}e^{-4\phi}$ and $\phi^{\prime\prime}$ from (\ref{ASMbeforeh}) and (\ref{ASMphipp}) into (\ref{ASMafterh}), we obtain
\be
\left(\phi^{\prime}+\frac{b^{\prime}}{b\,}\right)\left[
\left(\frac{b^{\prime}}{b\,}\right)^{\prime}+2\left(\frac{b^{\prime}}{b\,}\right)^{2}-\frac{2\,}{l^{2}}\right]=0\,.
\ee
We assume  $H\neq 0$ to get
\be
\left(\frac{b^{\prime}}{b\,}\right)^{\prime}+2\left(\frac{b^{\prime}}{b\,}\right)^{2}-\frac{2\,}{l^{2}}=0\quad~\Longleftrightarrow~\quad
\left(b^2\right)^{\prime\prime}=\frac{4}{l^2}b^2\,.
\ee
With the boundary condition $\phi(\eta_{0})=\phi_{0}$ and $a(\eta_{0})=a_{0}$, we have $b(\eta_{0})=e^{-\phi_{0}}a_{0}$ and thus with some constant $\alpha$,
\be
b^2(\eta)=\frac{e^{-2\phi_{0}}a_{0}^2}{2}\Big[(1+\alpha) e^{{2(\eta-\eta_{0})}/{l}}+(1-\alpha) e^{-{2(\eta-\eta_{0})}/{l\,}}\Big]\,.
\label{ASMb2}
\ee
We rewrite (\ref{ASMbeforeh})  by substituting this  expression of $b^2(\eta)$,
\be
b^4\phi^{\prime 2}+\frac{\hh^{2}}{4}e^{-4\phi}=\frac{3}{4}
\left[\left(b^2\right)^{\prime \,2}-\frac{4}{l^{2}}b^4\right]=
\frac{3a_{0}^4}{l^2e^{4\phi_{0}}}\left(\alpha^{2}-1\right)\,,
\ee
which  implies $\alpha^2\geq 1$ and, up to a certain sign factor $\sigma_{0}$ ($ {\sigma_{0}^2=1}$),
\be
\ba{rll}
\frac{\rd \eta}{b^{2}}&=&\scalebox{1.1}{$\frac{\sigma_{0}\rd\phi}{\sqrt{\frac{3}{l^2}\left(\alpha^{2}-1\right)e^{-4\phi_{0}}a_{0}^4-
\frac{\hh^{2}}{4}e^{-4\phi}}}$}\\
{}\!\!\!&\!\!\!\!\!\!=\!\!\!&\!\!\!\!\!\!\!\!
\frac{\sigma_{0} l e^{2\phi_{0}}}{2a_{0}^2\sqrt{3\left(\alpha^2-1\right)}}\,
\rd\ln\left[e^{2(\phi-\phi_{0})}{+\sqrt{e^{4(\phi-\phi_{0})}-\frac{\hh^2 l^2}{12a_{0}^{4}\left(\alpha^{2}-1\right)}}}\,
\right]\,.
\ea
\label{ASMeq1}
\ee
\begin{comment}
\be
\frac{\rd t}{b^{2}}=\frac{\sigma\rd\phi}{\sqrt{\frac{3}{l^2}\left(\alpha^{2}-1\right)e^{-4\phi_{0}}-
\frac{\hh^{2}}{4}e^{-4\phi}}}=
\frac{\sigma l e^{2\phi_{0}}}{2\sqrt{3\left(\alpha^2-1\right)}}\,\rd\left[\mathrm{cotanh}^{-1}\left(\frac{1}{\sqrt{1-\frac{\hh^{2}l^2}{12\left(\alpha^2-1\right)}e^{-4(\phi-\phi_{0})}}}\right)
\right]\,,
\label{ASMeq1}
\ee
\end{comment}

While so,   directly from (\ref{ASMb2}) we also get
\be
\ba{rll}
\frac{\rd \eta}{b^{2}}&=&\scalebox{1.2}{$\frac{2 e^{2\phi_{0}}a_{0}^{-2}\rd \eta}{(\alpha+1) e^{{2(\eta-\eta_{0})}/{l}}-(\alpha-1) e^{-{2(\eta-\eta_{0})}/{l}}}$}\\
{}&=&\scalebox{1.2}{$\frac{le^{2\phi_{0}}}{2a_{0}^2\sqrt{\alpha+1}\sqrt{\alpha-1}}\,\rmd\ln\left[\frac{\sqrt{\frac{\alpha+1}{\alpha-1}}e^{{2(\eta-\eta_{0})}/{l}}-1}{\sqrt{\frac{\alpha+1}{\alpha-1}}e^{{2(\eta-\eta_{0})}/{l}}+1}\right]\,.$}
\ea
\label{ASMeq2}
\ee
Matching  (\ref{ASMeq1}) with (\ref{ASMeq2}), we  obtain
\be
\ba{rll}
e^{2\phi(\eta)}&=&\frac{l\left|\hh\right| e^{2\phi_{0}}}{4a_{0}^{2}\sqrt{3\left(\alpha^2-1\right)}}\left[\beta\left\{\frac{\sqrt{\frac{\alpha+1}{\alpha-1}}e^{{2(\eta-\eta_{0})}/{l}}-1}{\sqrt{\frac{\alpha+1}{\alpha-1}}e^{{2(\eta-\eta_{0})}/{l}}+1}\right\}^{\sqrt{3}}\right.\\
{}&{}&\left.
~\qquad+\beta^{-1}\left\{\frac{\sqrt{\frac{\alpha+1}{\alpha-1}}e^{{2(\eta-\eta_{0})}/{l}}-1}{\sqrt{\frac{\alpha+1}{\alpha-1}}e^{{2(\eta-\eta_{0})}/{l}}+1}\right\}^{-\sqrt{3}}\,
\right]\,,
\ea
\ee
where $\beta$ is an  integral constant  that has effectively absorbed  the sign factor, $\frac{\sigma_{0}\sqrt{\alpha+1}\sqrt{\alpha-1}}{\sqrt{\alpha^2-1}}$.

Imposing  $\phi(\eta_{0})=\phi_{0}$,  we   fix $\beta$ with another sign factor $\sigma$ ($\sigma^2=1$), 
\be
\beta=\!\small{\left[\frac{\sqrt{12\left(\alpha^2-1\right)}}{l\left|\hh\right| a_{0}^{-2}}+\sigma\sqrt{\frac{12\left(\alpha^2-1\right)}{(l\hh )^2a_{0}^{-4}}-1\,}\,\right]\!\left(\frac{\sqrt{\frac{\alpha+1}{\alpha-1}}-1}{\sqrt{\frac{\alpha+1}{\alpha-1}}+1}\right)^{\!-\sqrt{3}}\!\!.}
\ee
Now, if we set
\be
{\zeta}\equiv\ln\sqrt{\frac{\alpha+1}{\alpha-1}}\,,
\ee
we arrive at our final forms of the dilaton~(\ref{analyticphi}), 
\be
\ba{l}
e^{2\phi(\eta)-2\phi_{0}}=\frac{1}{2}{\textstyle{\left(1{+\sigma}\sqrt{1-\frac{(\hh l\sinh\zeta)^2}{12a_{0}^{4}}}\right)}}\!\left\{
\frac{\tanh\left(\frac{\eta-\eta_{0}}{l}+\frac{\zeta}{2}\right)}{\tanh\frac{\zeta}{2}}\right\}^{\sqrt{3}}\\
\qquad
+\frac{1}{2}{\textstyle{\left(1{-\sigma}\sqrt{1-\frac{(\hh l\sinh\zeta)^2}{12a_{0}^{4}}}\right)}}\!\left\{
\frac{\tanh\left(\frac{\eta-\eta_{0}}{l}+\frac{\zeta}{2}\right)}{\tanh\frac{\zeta}{2}}\right\}^{-\sqrt{3}}\,,
\ea
\label{Ae2p}
\ee
and, in terms of $\zeta$,  the scale factor~(\ref{SF}),
\be
a^2(\eta)=e^{2\phi(\eta)}b^{2}(\eta)
=\frac{a_{0}^{2}e^{2\phi(\eta)}\sinh\left({2(\eta-\eta_{0})}/{l}+\zeta\right)}{e^{2\phi_{0}}\sinh\zeta}\,.
\label{Ab2}
\ee
We proceed to determine $\zeta$ in terms of $H_{0}$: From
\be
\ba{rll}
H_{0}&=&\left.\frac{1}{2e^{3\phi}b^3}\left[b^2(\eta)e^{2\phi(\eta)}\right]^{\prime}\right|_{\eta=\eta_{0}}\\
{}&=&\frac{1}{2a_{0}l\sinh\zeta}\left[2\cosh\zeta+
{\sigma\sqrt{12-a_{0}^{-4}\left(\hh l\sinh\zeta\right)^2}}
\right]\,,
\ea
\label{AH0far}
\ee
we acquire,  with two sign factors, $\sigma_{1},\sigma_{2}$ ($\sigma_{1}^2=1=\sigma_{2}^2$),
\be
\sinh\zeta=\textstyle{\frac{\sigma_{1}2\sqrt{2}}{\sqrt{
4{+l^2}\!\left(8a_{0}^2H_{0}^2+\frac{\hh^2}{a_{0}^{4}}\right)\!+2\sigma_{2}a_{0} H_{0}l\sqrt{24+2l^2
\left(6a_{0}^2H_{0}^2+\frac{\hh^2}{a_{0}^{4}}\right)\,}}}\,.}
\label{Asinzetapre}
\ee
Substituting this back into the last expression in (\ref{AH0far}), assuming $H_{0}>0$ and $l>0$,  we arrive 
\be
\ba{lll}
H_{0}&\!\!=&\!\!\frac{\sigma_{1}}{2\sqrt{2}a_{0} l}\left[\,
\sqrt{12+l^2\left(6a_{0}^2H_{0}^2+\hh^2a_{0}^{-4}\right)}+\sigma_{2}\sqrt{2}a_{0} H_{0}l\right.\\
{}&\!\!\!&\!\!\!\!\!\!\left.+
\sigma \left.\sqrt{12+l^2\left(6a_{0}^2H_{0}^2+\hh^2a_{0}^{-4}\right)}+\sigma_{2}3\sqrt{2}a_{0} H_{0}l\,\right|\,\right]\,.
\ea
\ee
Thus, to meet this algebraic identity, we must have either 
{case I} where $\sigma_{2}=1$ and hence $\sigma=-1$, $\sigma_{1}=-1$;\\ or 
{case II} where $\sigma_{2}=-1$ and 
\begin{itemize}
\item[--] if $12+\frac{\hh^2l^2}{a_{0}^{4}}< 12a_{0}^2 H_{0}^2 l^2$, then $\sigma=1$ and $\sigma_{1}=1$\,,
\item[--] if $12+\frac{\hh^2l^2}{a_{0}^{4}}> 12a_{0}^2 H_{0}^2 l^2$, then  $\sigma=-1$ and $\sigma_{1}=1$\,.
\end{itemize}
Consequently, we  always have $\sigma_{2}=-\sigma_{1}$ and
\be
\ba{l}
\sinh\zeta=\frac{-2\sigma_{2}\sqrt{2}}{\sqrt{
4+l^2\left(8a_{0}^2H_{0}^2+\frac{\hh^2}{a_{0}^{4}}\right)+2\sigma_{2}a_{0} H_{0}l\sqrt{24+2l^2
\left(6a_{0}^2H_{0}^2+\frac{\hh^2}{a_{0}^{4}}\right)\,}}}\,,\\
\tanh\zeta=\frac{-2\sigma_{2}\sqrt{2}}{\sqrt{12+l^2\left(6a_{0}^2H_{0}^2+\frac{\hh^2}{a_{0}^{4}}\right)}+\sigma_{2}\sqrt{2} a_{0}H_{0}l}\,.
\ea
\ee
It is straightforward to check that indeed (\ref{Ae2p}) and (\ref{Ab2}) satisfy  (\ref{ASMreduced2}): for the former of (\ref{ASMreduced2}), the squares of each side of the equality match. 

In order to  determine the sign factors completely, we further compute
\be
\left.\phi(\eta)^{\prime}\right|_{\eta=\eta_{0}}=\sigma\frac{\sqrt{12-a_{0}^{-4}\left(\hh l\sinh\zeta\right)^2}}{2l\sinh\zeta}\,,
\ee
and
\be
\ba{rll}
\left.\frac{2\phi(\eta)^{\prime}}{N(\eta)}-3H(\eta)\right|_{\eta=\eta_{0}}\!\!
&=&\!
-\left[\frac{6\cosh\zeta+\sigma\sqrt{12-a_{0}^{-4}\left(\hh l\sinh\zeta\right)^2}}{2a_{0}l\sinh\zeta}\right]\\
{}&=&\sigma_{2} \sqrt{ 3H_{0}^{2}+\frac{\hh^2}{2a_{0}^{6}}+\frac{6}{a_{0}^{2}l^{2}}\,}\,.
\ea
\ee
This shows that $\sigma_{2}$ must coincide with the upper/lower $\pm$ sign in (\ref{ASMreduced2}).   Consequently, (\ref{Asinzetapre}) becomes
\be
\!\sinh\zeta=\! \scalebox{1.1}{$\frac{\mp2\sqrt{2}}{\sqrt{
4{+l^2}\big(8a_{0}^2H_{0}^2{+\frac{\hh^2}{a_{0}^{4}}}\big){\pm 2} a_{0}H_{0}l\sqrt{24{+2l^2}
\big(6a_{0}^2H_{0}^2{+\frac{\hh^2}{a_{0}^{4}}}\big)}}}$}\,,
\ee
and $\sigma$ is fixed as
\be
\sigma=\,-\,\frac{\sqrt{12+l^2\left(6a_{0}^2H_{0}^2+\hh^2a_{0}^{-4}\right)}\pm 3\sqrt{2}la_{0}H_{0}}{\left|\sqrt{12+l^2\left(6a_{0}^2H_{0}^2+\hh^2a_{0}^{-4}\right)}\pm 3\sqrt{2}la_{0}H_{0}\right|}\,.
\ee
With (\ref{defO}), these lead to (\ref{szs}).   Note the existing  freedom to change $a_{0},\eta_{0},\phi_{0},\zeta$, and $\sigma$ while keeping $e^{\phi(\eta)}$~(\ref{Ae2p}), $a(\eta)$~(\ref{Ab2}), and consequently   $\zeta{-2\eta_{0}}/l$ all fixed.
\vspace{9pt}

\subsection*{Acceleration}
From the analytic solution, it is straightforward to compute  the acceleration, 
\be
\ba{l}
{\left.\frac{1}{a}\left(\frac{\rmd~}{N\rmd\eta}\right)^2a(\eta)\right|_{\eta=\eta_{0}}
=\left.\frac{\,\rmd H(\eta)}{N\rmd\eta}+H(\eta)^2\right|_{\eta=\eta_{0}}}\\
{\quad=
\frac{\left(\hh la_{0}^{-2}\sinh\zeta\right)^2-4-2\sigma\cosh\zeta\sqrt{12-
\left(\hh la_{0}^{-2}\sinh\zeta\right)^2} }{2\left(a_{0}l\sinh\zeta\right)^2}}\\
{\quad=H_{0}^{2}+\frac{2}{a_{0}^2l^2}+\frac{\hh^{2}}{2a_{0}^{6}}\pm H_{0}\sqrt{3H_{0}^{2}+
\frac{6}{a_{0}^2l^2}+\frac{\hh^{2}}{2a_{0}^{6}}}\,.}
\ea
\label{Afslines}
\ee
The  expression on the middle line implies that  the acceleration  is  positive for $(\hh la_{0}^{-2}\sinh\zeta)^2\leq 12$ and $\sigma=-1$, since
\be
\ba{c}
\left(\frac{\hh l}{a_{0}^{2}}\sinh\zeta\right)^2-4+2\cosh\zeta\sqrt{12-
\left(\frac{\hh l}{a_{0}^{2}}\sinh\zeta\right)^2}\\\geq \left(\frac{\hh l}{a_{0}^{2}}\sinh\zeta\right)^2\!-4+2\sqrt{12-
\left(\frac{\hh l}{a_{0}^{2}}\sinh\zeta\right)^2}\,\geq\,4(\sqrt{3}-1)\,.
\ea
\ee
Additionally, the last line in (\ref{Afslines}) shows that it is positive,  always  for the upper  $+$ sign, or in the case of the lower  $-$  sign  if  $a_{0}H_{0}l$ satisfies an inequality,
\be
\ba{rll}
a_{0}H_{0}l&<&\textstyle{\frac{1}{2\sqrt{2}}\sqrt{\frac{(\hh l)^2}{a_0^4}-4 +\sqrt{144+56\frac{(\hh l)^2}{a_0^4}+9\frac{(\hh l)^4}{a_0^8}}}}\\
{}&{}&~\simeq
1+\frac{5}{24}\big(\frac{\hh l}{a_{0}^2}\big)^2-\frac{43}{3456}\big(\frac{\hh l}{a_{0}^2}\big)^4\,.
\ea
\label{accinequalpre}
\ee
This is equivalent, in terms of $\Omega_{k}$ and $\Omega_{\hh}$, to 
\be
{\Omega_{k}\,>\,\frac{1}{4}-3\Omega_{\hh}+\frac{3}{4}\sqrt{1-\frac{16}{3}\Omega_{\hh}}~\simeq\, 1-5\Omega_{\hh}\,.}
\ee

Finally, it becomes  possible to  solve for $\hh la_{0}^{-2}$ in terms of $q_{0}$ and $a_{0}H_{0}l$, 
\be
\scalebox{0.99}{$\!\!\left(\frac{\hh l}{a_{0}^{2}}\right)^{\!2}\!\!=$}\scalebox{0.88}{$ \,-4-(1{+2q_{0}})(a_{0}H_{0}l)^2\mp\!\sqrt{16(a_{0}H_{0}l)^2+(9{-4q_{0}})(a_{0}H_{0}l)^4}\,,$}
\ee
and to take  $\big\{H_{0},q_{0},l\big\}$  as  independent  parameters alternative to $\big\{H_{0},\hh,l\big\}$, if desired.\newpage
%\vspace{9pt}%~\\

\begin{comment}
\section*{Einstein Frame}
Converted to the Einstein frame, $g^{\rm{E}}_{\mu\nu}=e^{-2\phi}g_{\mu\nu}$, we have
\be
\ba{lll}
N_{\rm{E}}=e^{-\phi}N\,,\quad&\quad a_{\rm{E}}=e^{-\phi}a\,,\quad&\quad H_{\rm{E}}=\frac{\,\rd \ln a_{\rm{E}}}{N_{\rm{E}}\rd t}\,.
\ea
\label{AEN}
\ee
From (\ref{OFE}) and (\ref{conservation}),  the deceleration parameter is given by
\be
\ba{l}
q_{\rm{E}}=-
\frac{1}{H_{\rm{E}}^{2}a_{\rm{E}}}\Big(\frac{\rmd~}{N_{\rm{E}}\rmd t}\Big)^{\!2}a_{\rm{E}}\\
\quad
=\frac{1}{6H^{2}_{\rm{E}}}\left[\left(\rhonew+3\pnew\right)e^{4\phi}+\frac{\hh^2e^{-4\phi}}{a_{\rm{E}}^{6}}+\left(\frac{2\rmd \phi}{N_{\rm{E}}\rmd t}\right)^{2}-\To e^{2\phi}\right]\,.
\ea
\label{AEq}
\ee
Clearly, unless  the strong energy condition $\rhonew+3\pnew>0$ is violated or $\To>0$ is dominant, $q_{\rm{E}}$ is generically  positive  and thus the deceleration is natural in the Einstein frame.   It is remarkable that, depending on the frame, \textit{i.e.}~string or  Einstein,   the positive pressure $\pnew>0$   accelerates or decelerates  the expansion of the Universe. \\
\vspace{9pt}
\end{comment}
\section*{Supplemental  Bayesian Inferences}
\renewcommand\thesection{SM}
\renewcommand{\thefigure}{SM\,\arabic{figure}}
\setlength{\jot}{9pt}                 % spacing btwn the rows of an eqnarray
\renewcommand{\arraystretch}{3.2} 
\setcounter{equation}{0}
\setcounter{figure}{0}
Below we present more Bayesian Inference (BI) results using different combinations of free parameters. As done for FIGs.\,\ref{FIGBI6p},\,\ref{FIGBI6pwfree},\,\ref{FIGBI2p}, we consistently use $100$ walkers and $10^6$ steps to run the BI on a supercomputer. %We notice that some parameters, such as the curvature density, have a strong influence on the others, creating an approximate linear dependence among them. This makes it difficult to estimate the uncertainty of those parameters accurately. However, the overall BI results are consistent and coherent with each other. 
%\newpage
\begin{figure}[H]
\begin{center}
\includegraphics[width=0.96\linewidth]{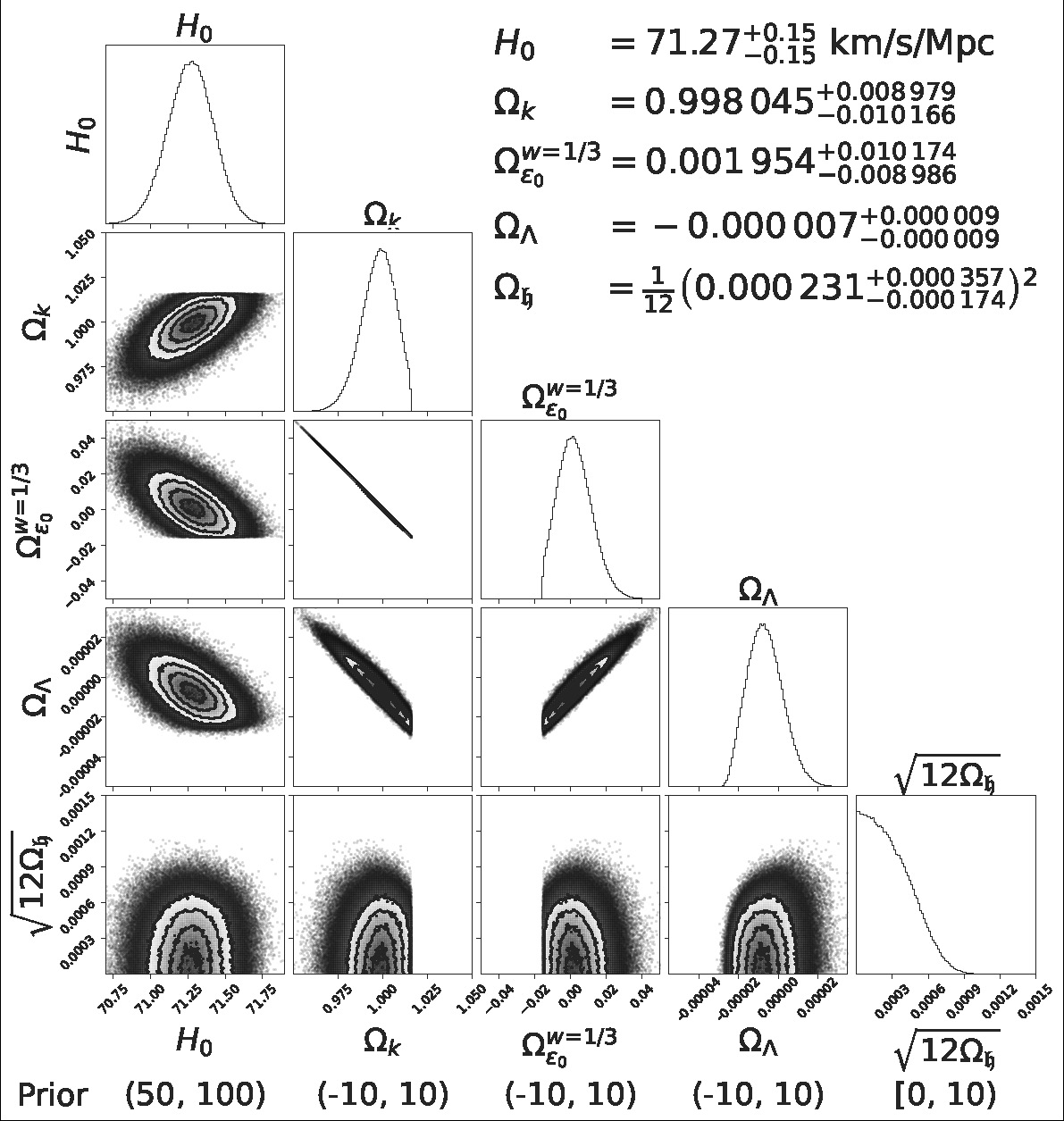}
\caption{{{Bayesian Inference with Five  Parameters I:  \\ Single  Fluid with ${w=1/3}$ Fixed.}} \label{SMFIGBI5p}}
~\\
~\\
\includegraphics[width=0.96\linewidth]{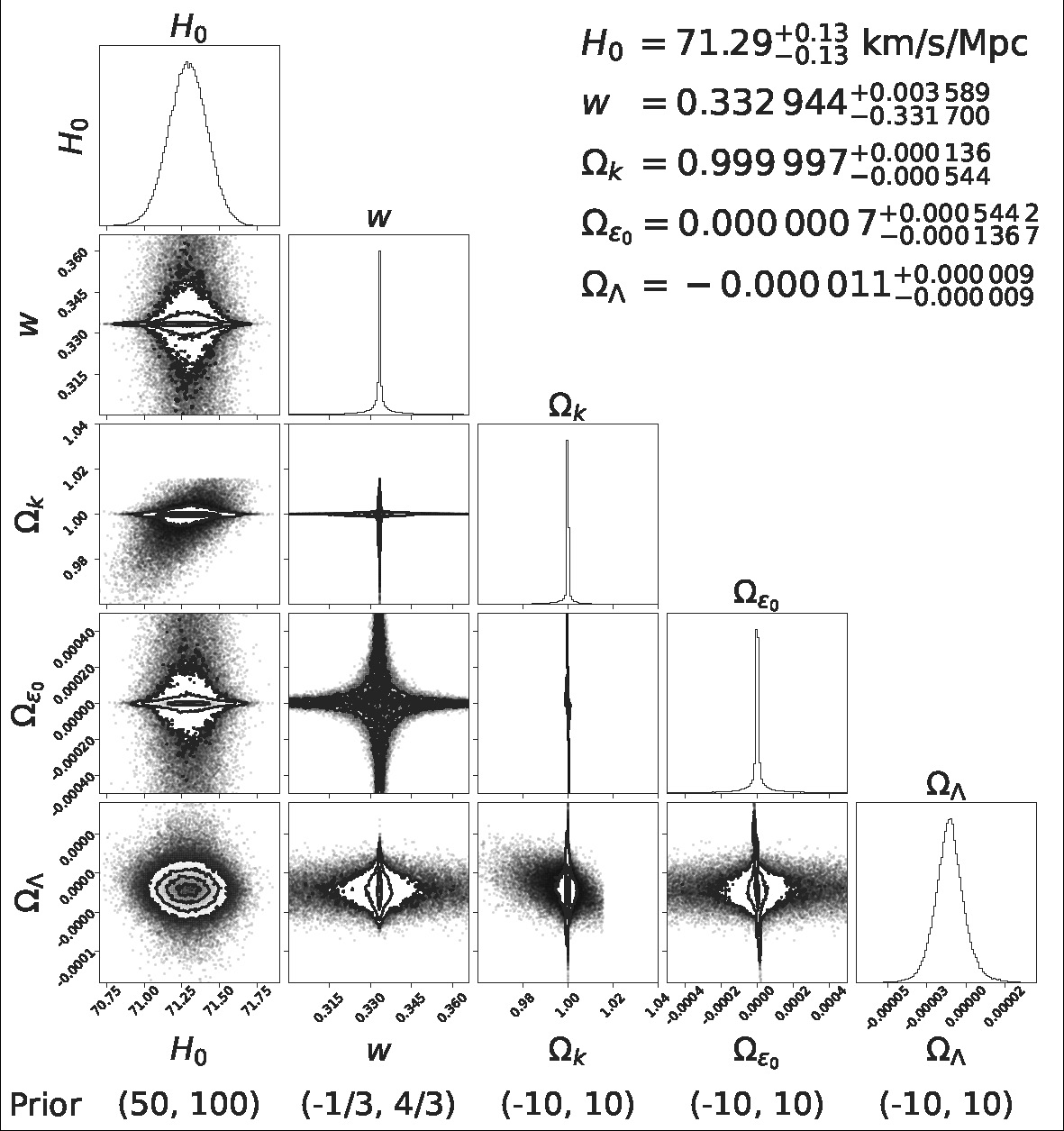}
\caption{{{Bayesian Inference with Five  Parameters II: \\Another indication of a possible bias toward  $w=1/3$.}} \label{SMFIGBI5pwfree}}
\end{center}
\end{figure}
\newpage
\noindent We notice that some parameters, such as the curvature density, have a strong influence on the others, creating an approximate linear dependence among them. This makes it difficult to estimate the uncertainty of those parameters accurately. However, the overall BI results are consistent and coherent with each other. 
\begin{figure}[H]
\begin{center}
\includegraphics[width=0.96\linewidth]{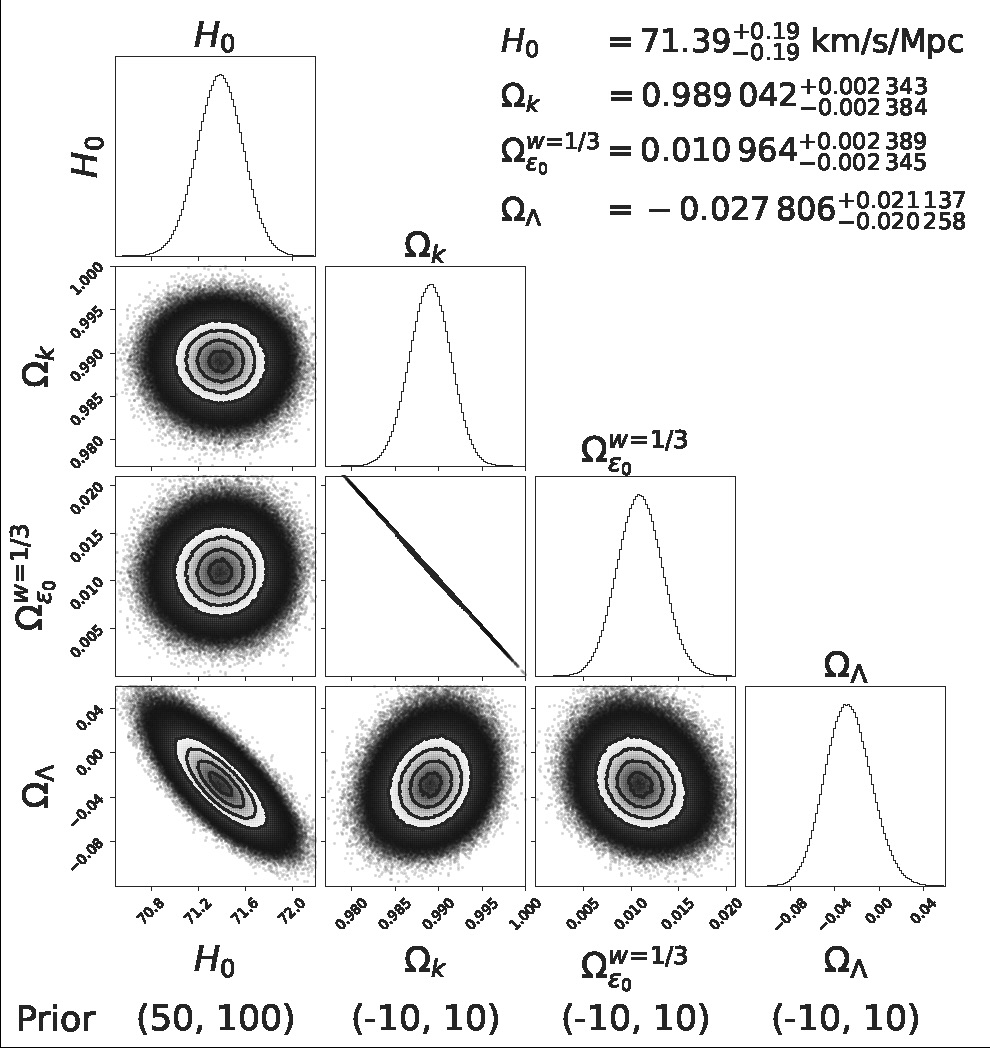}
\caption{{{Bayesian Inference  with Four  Parameters I:\\ Negative Cosmological Constant, $\Lambda<0$. }} \label{SMFIGBI4pLambda}}
~\\
\includegraphics[width=0.96\linewidth]{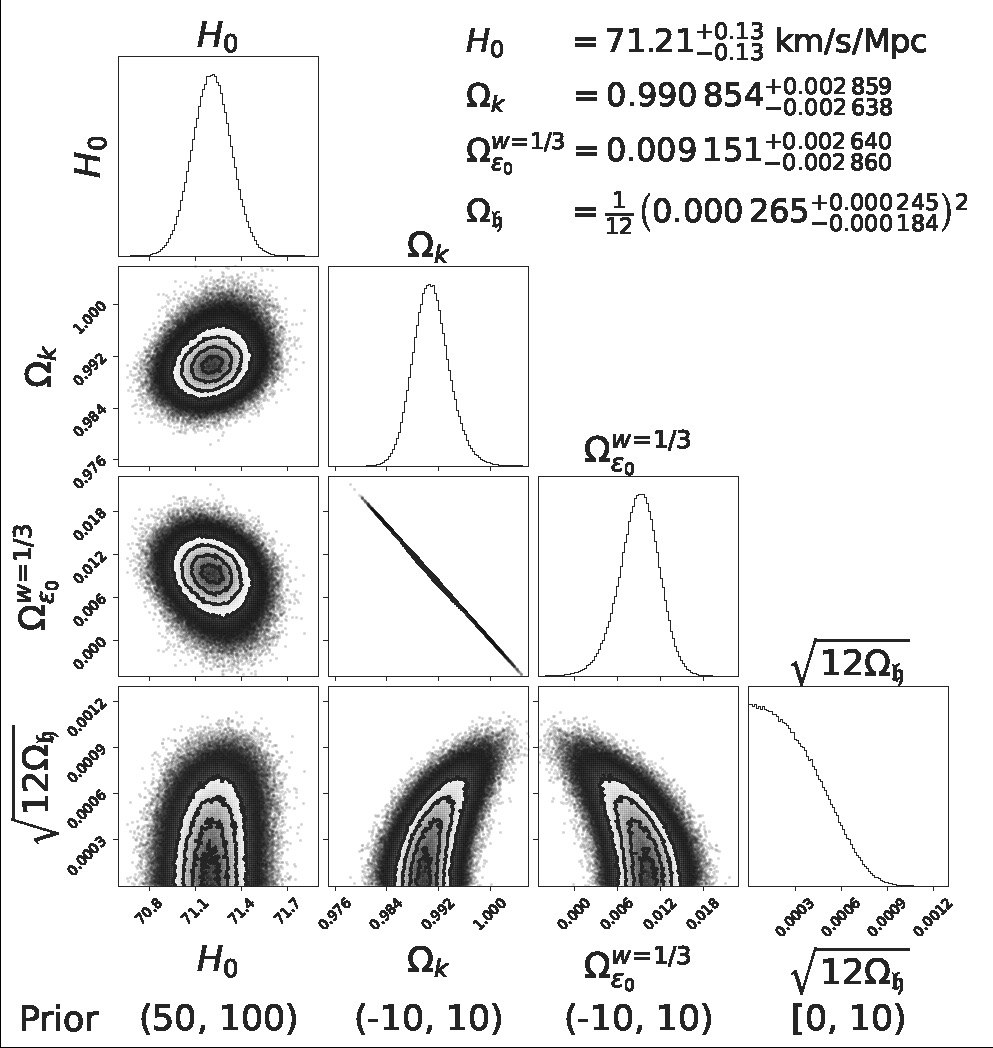}
\caption{{{Bayesian Inference  with Four  Parameters II:\\ Non-trivial $H$-flux, $\hh\neq0$. }} \label{SMFIGBI4p}}
\end{center}
\end{figure}
\newpage
\begin{figure}[H]
\begin{center}
\includegraphics[width=0.96\linewidth]{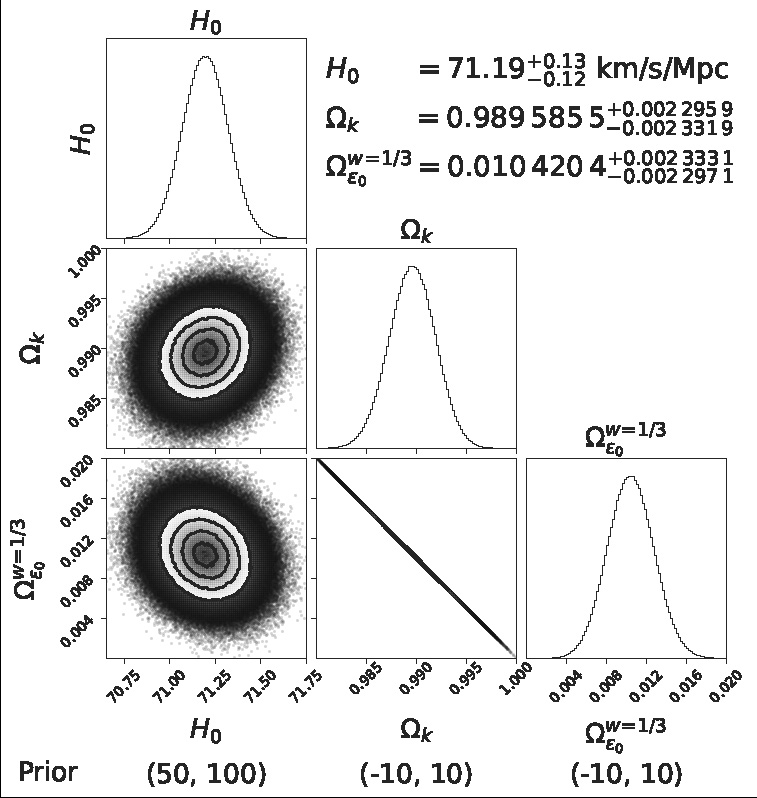}
\caption{{{Bayesian Inference with  Significant Three  Parameters.}} \label{SMFIGBI3p}}
~\\
~\\
~\\
\includegraphics[width=0.96\linewidth]{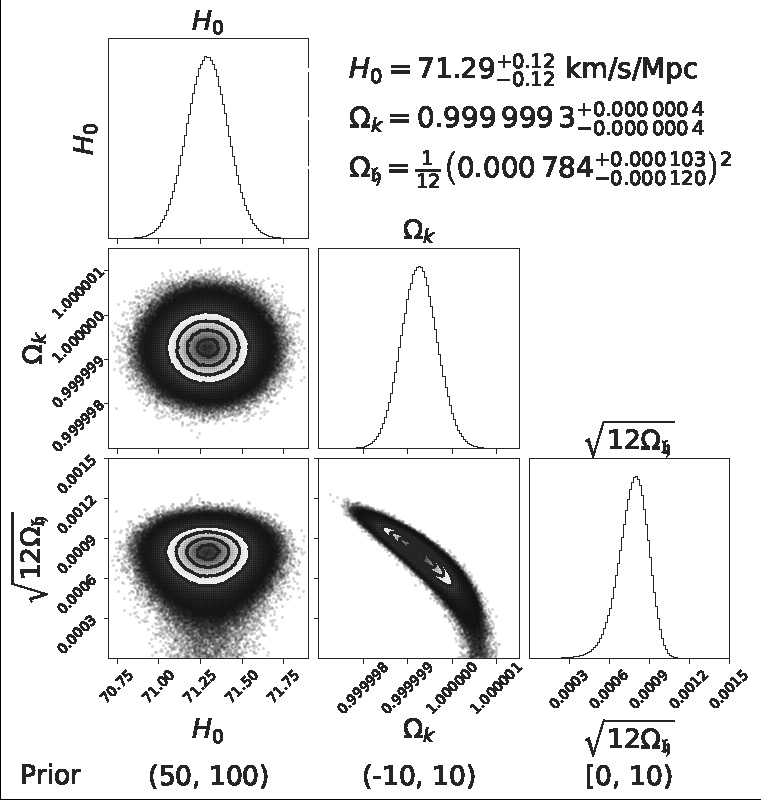}
\caption{{{Bayesian Inference for  Pure Vacuum Geometry with Three Parameters.}} \label{SMFIGBIVAC3p}}
\end{center}
\end{figure}
%\newpage
\section*{Hubble  \& Deceleration Parameters Elongated up to the Bouncing  Points of  ${H(z)=0}$ \&  ${q(z)=-\infty}$}
\begin{figure}[H]
\begin{center}
\includegraphics[width=0.9\linewidth]{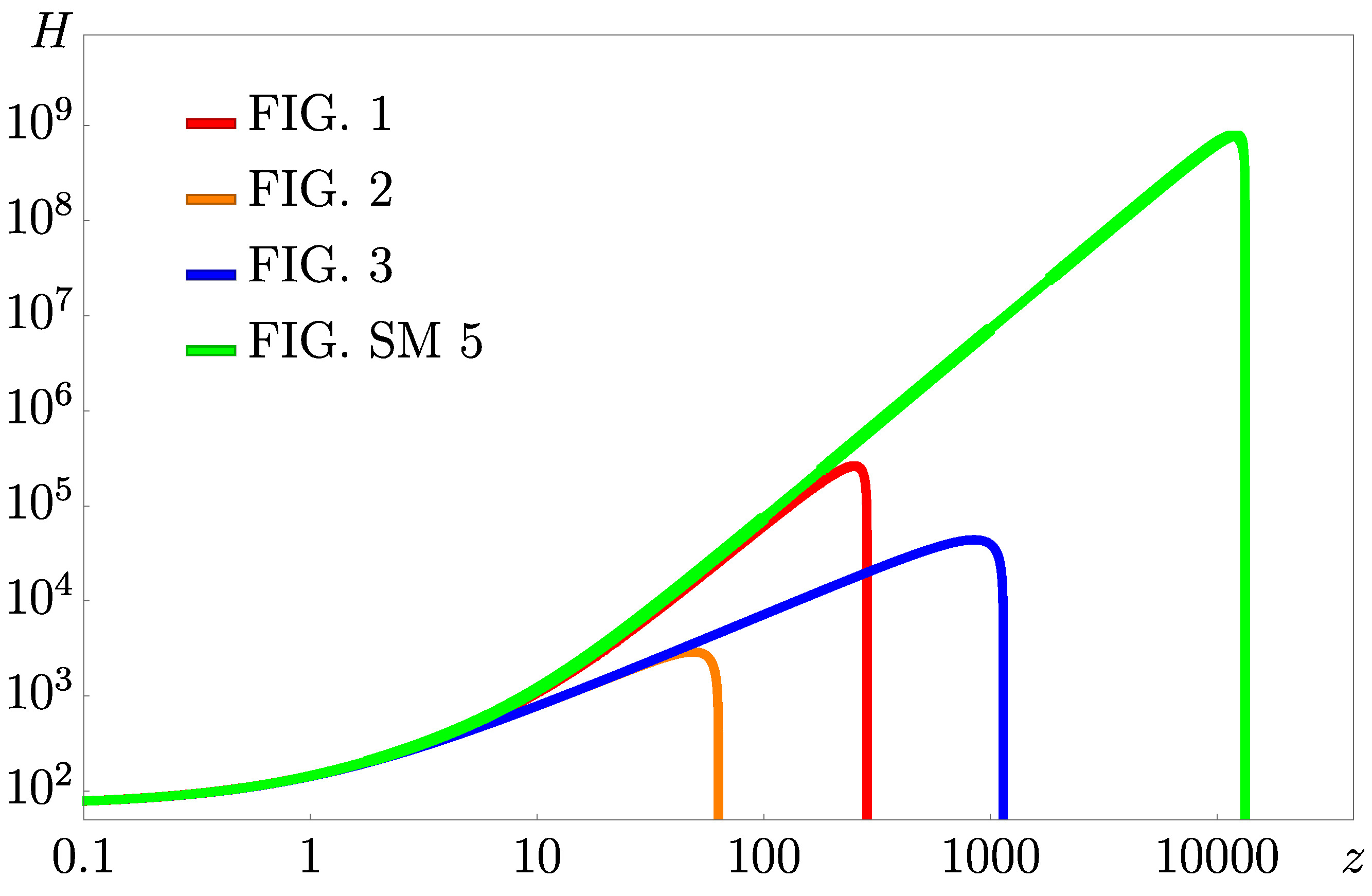}
\caption{{{Elongated Hubble Parameter based on the  BI results of FIGs.\,\ref{FIGBI6p},  \ref{FIGBI6pwfree}, \ref{FIGBI2p}, \& \ref{SMFIGBI3p}.  }} \label{SMFIGHubble}}
~\\
%~\\
%~\\
\includegraphics[width=0.9\linewidth]{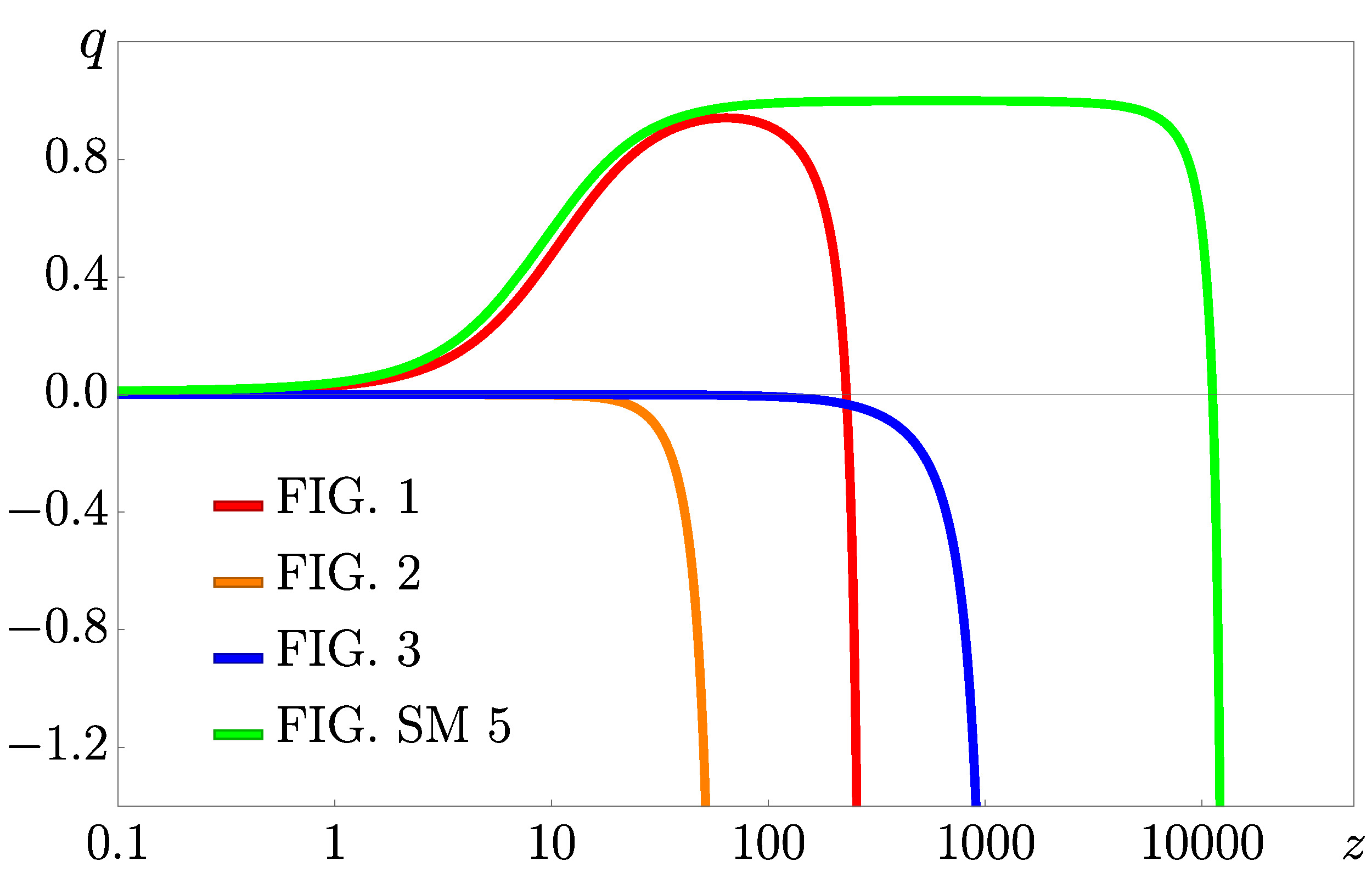}
\caption{{{Elongated Deceleration Parameter based on the  BI results of   FIGs.\,\ref{FIGBI6p},  \ref{FIGBI6pwfree}, \ref{FIGBI2p}, \&  \ref{SMFIGBI3p}. 
 To be consistent with the analytic vacuum geometry~(\ref{FutureInfty}), 
our BI fitting results indicate that the present time acceleration is  marginal: $q_{0}$ amounts to $0.0075\pm 0.0128$,   $- 10^{-5}$, or  $- 6{\times 10^{-7}}$ for FIGs.\,\ref{FIGBI6p},\,\ref{FIGBI6pwfree},\,\ref{FIGBI2p} respectively.  Nevertheless,  $q \rightarrow -\infty$  when  bouncing~(\ref{bouncing}). \label{SMFIGq}}}}
%\end{center}
%\end{figure} 
~\\
%\section*{Bouncing Scale Factor with respect to the\\ Cosmic Proper Time }
%\begin{figure}[H]
%\begin{center}
\includegraphics[width=0.9\linewidth]{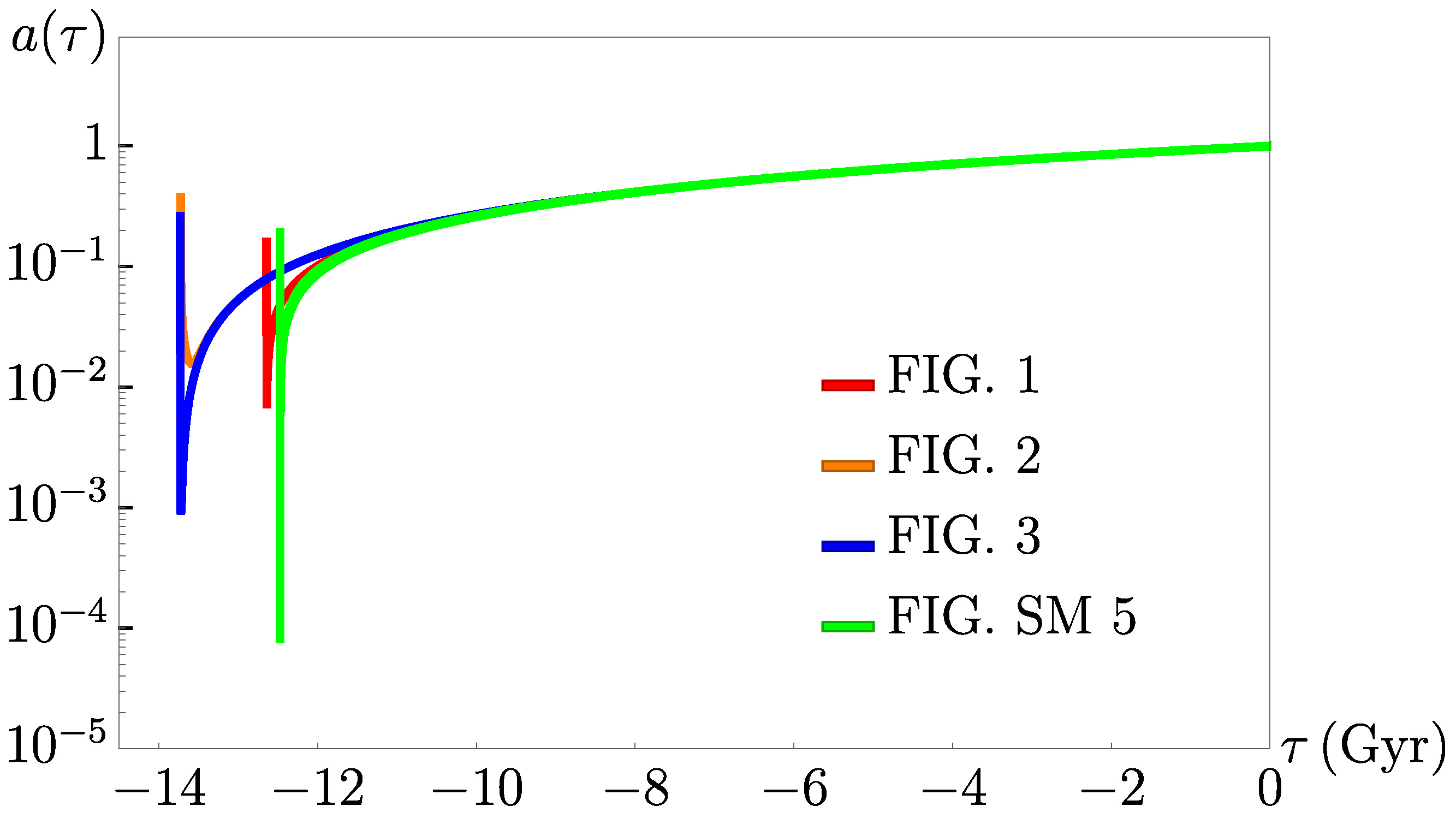}
\caption{{{Bouncing scale factor of the open Universe with respect to the cosmic proper time, $\tau=\int_{0}^{t} {\rm{d}}t^{\prime}\,N(t^{\prime})$, based on the BI results of FIGs.\,\ref{FIGBI6p}, \ref{FIGBI6pwfree}, \ref{FIGBI2p}, \& \ref{SMFIGBI3p}. The bouncing occurs \big\{12.64,\, 13.58,\, 13.72,\, 12.47\big\} gigayears ago  respectively, and the minimum scale factor $\min[a]=1/(1+\max[z])$ is given by the maximum  redshift,  \big\{286.05,\, 63.18,\, 1141.36,\,   13323.86\big\}. It is remarkable that, though $\max[z]$ varies significantly, the bouncing time is close to the "age" of the flat Universe estimated in the $\Lambda$CDM model.}}} \label{SMFIGHubble}
\end{center}
\end{figure}

%\newpage

%HL and LY contributed equally to this work. Correspondence should be addressed to JHP:    \url{park@sogang.ac.kr}  

\end{document}